\documentclass[journal]{IEEEtran}
\usepackage{amsthm,amssymb}
\usepackage[english]{babel}
\usepackage[para]{footmisc}
\usepackage{cite}
\usepackage{graphicx}
\usepackage{mathtools}
\usepackage{amsmath}
\usepackage{caption}
\usepackage{epstopdf}
\usepackage{tabularx}
\usepackage[section]{placeins}
\usepackage{cuted}

\usepackage{setspace}
\usepackage{stfloats}
\usepackage{multirow}
\usepackage{float}
\usepackage{algorithm}
\usepackage{algorithmic}
\usepackage[flushleft]{threeparttable}
\usepackage{xcolor}
\usepackage{enumerate}
\usepackage[colorlinks,linkcolor=red,citecolor=blue]{hyperref}
\usepackage{subfigure}

\usepackage{color}
\usepackage{tabularx}
\newcommand{\tabincell}[2]{\begin{tabular}{@{}#1@{}}#2\end{tabular}} 

\usepackage{titletoc}
\usepackage{mathrsfs}
\renewcommand\thesubsubsection{\Alph{subsubsection}}

\ifCLASSINFOpdf

\else

\fi

\begin{document}

\title{Physical Layer Security for UAV Communications in 5G and Beyond Networks}

\author{
\IEEEauthorblockN{Jue Wang, {\em Member, IEEE}, Xuanxuan Wang, Ruifeng Gao, {\em Member, IEEE}, Chengleyang Lei, {\em Student Member, IEEE}, Wei Feng, {\em Senior Member, IEEE}, Ning Ge, {\em Member, IEEE}, Shi Jin, {\em Senior Member, IEEE}, \\and Tony Q. S. Quek, {\em Fellow, IEEE}
}
\thanks{Jue Wang and Ruifeng Gao are with School of Information Science and Technology, Nantong University, Nantong 226019, China. Email: \{wangjue, grf\}@ntu.edu.cn}
\thanks{Xuanxuan Wang ({\em Co-first author}), Chengleyang Lei, Wei Feng ({\em Corresponding author}), and Ning Ge are with Beijing National Research Center for Information Science and Technology, Department of Electronic Engineering, Tsinghua University, Beijing 100084, China. X. Wang is also with the China Academy of Electronics and Information
Technology, Beijing 100041, China. Email: wangxuanxuan@mail.tsinghua.edu.cn, lcly17@mails.tsinghua.edu.cn, \{fengwei, gening\}@tsinghua.edu.cn}
\thanks{Shi Jin is with National Mobile Communications Research Laboratory, Southeast University, Nanjing 210096, China. Email: jinshi@seu.edu.cn}
\thanks{Tony Q. S. Quek is with Information Systems Technology and Design Pillar,
Singapore University of Technology and Design, Singapore 487372. Email: tonyquek@sutd.edu.sg}
}

\maketitle

\begin{abstract}
Due to its high mobility and flexible deployment, unmanned aerial vehicle (UAV) is drawing unprecedented interest in both military and civil applications to enable agile wireless communications and provide ubiquitous connectivity.
Mainly operating in an open environment, UAV communications can benefit from dominant line-of-sight links; however, it on the other hand renders the UAVs more vulnerable to malicious eavesdropping or jamming attacks.
Recently, physical layer security (PLS), which exploits the inherent randomness of the wireless channels for secure communications, has been introduced to UAV systems as an important complement to the conventional cryptography-based approaches. In this paper, a comprehensive survey on the current achievements of the UAV-aided wireless communications is conducted from the PLS perspective. We first introduce the basic concepts of UAV communications including the typical static/mobile deployment scenarios, the unique characteristics of  
air-to-ground channels, as well as various roles that a UAV may act when PLS is concerned. Then, we introduce the widely used secrecy performance metrics and start by reviewing the secrecy performance analysis and enhancing techniques for statically deployed UAV systems, and extend the discussion to a more general scenario where the UAVs' mobility is further exploited. For both cases, respectively, we summarize the commonly adopted methodologies in the corresponding analysis and design, then describe important works in the literature in detail. Finally, potential research directions and challenges are discussed to provide an outlook for future works in the area of UAV-PLS in 5G and beyond networks.
\end{abstract}


\IEEEpeerreviewmaketitle

\section{Introduction}

Current achievements of the fifth generation (5G) technologies have provided significant advances in the conventional communication scenarios \cite{Andrews_2014JSAC, Wang-ICM2014}. However, it is still challenging to support the communication demand in the remote areas where the conventional communication infrastructure cannot reach (e.g., on the ocean) \cite{Saarnisaari_2020,Li_WC_2020}, or in the unexpected or emergency situations such as mass crowding or infrastructure malfunctioning due to natural disasters. Besides, the explosively increasing Internet-of-Things devices and applications also require new coverage enhancing techniques. To handle these issues, an intelligent architecture with the aid of unmanned aerial vehicles (UAVs)
has been considered as a promising new paradigm \cite{Erdelj_PC17,Luo_VTCSpring2015,Wan_2018Access,Motlagh_ICM2017}.
Owing to the merits of high mobility and flexible deployment, UAV communications have been widely invoked for civilian, commercial, and military applications in a wide way \cite{Zeng_Magazine_UAV124,Li_IOTJ_UAV47,Nokia_2018,Sekander_2018ICM,Zeng_Pro_2019}.
A series of excellent survey works can be found in the literature, to help the researchers establish a comprehensive understanding of the network architecture and techniques, as well as opportunities and challenges, that appeared in the current and future UAV communication systems \cite{UAVsurvey2015,Hayat_survey16,Fotouhi_survey19,Mozaffari_survey19}.

While providing promising merits, the UAV communication may suffer more from malicious attacks. One the one hand, the high flying altitude of UAV generally leads to strong line-of-sight (LoS) links between UAV and ground nodes, or among the aerial nodes. A dominant LoS air-to-ground (A2G) or air-to-air (A2A) link is generally desirable for legitimate communications, however, it also poses potential advantages for adversaries. An attacker could take advantage of the LoS propagation to enhance its eavesdropping quality, or improve its jamming efficiency. On the other hand, 
UAV may work in a shared frequency band with other systems in some cases. For example, most current UAV communication systems adopt the Industrial Scientific Medical (ISM) band defined by ITU-R (e.g., 2.4 GHz). The large amount of users, which operate in the same band but may come from different systems, will raise interference (unintended jamming) and privacy issues.
For these reasons, communication secrecy in UAV systems have attracted increasing research attentions, not only from the academia but also from the industry \cite{Welch_16}.

Unlike the terrestrial communication infrastructures (e.g., terrestrial base stations (BSs)), UAVs usually have more stringent constraints on their on-board energy and computational capability. Besides, due to its high mobility, the topology of a UAV communication network could be highly dynamic. For these reasons, applying the upper-layer cryptography-based schemes for secure UAV communications might have practical limitation. Physical layer security (PLS), on the other hand, has been recognized as a promising way to realize information theoretic secure transmissions with low computational complexity\cite{Mukherjee_ICST2014,Wu_JSAC18}. The application of PLS techniques to UAV communications has attracted dramatically increasing research interests since the past few years. It should be noted that the conventional PLS techniques in general cannot be directly applied to the UAV scenario, where some important new issues that need to be specifically taken into account include:

\begin{itemize}
\item {\em New design dimension:} The mobility of UAV raises a new dimension for optimization, that is, the position or trajectory of UAV. As a matter of fact, position optimization and trajectory planning constitute a large portion of the related research works on UAV-PLS.

\item{\em New channel/network characteristics:} The A2G/A2A channels have different characteristics compared with the terrestrial channels; the distribution of aerial nodes may have new forms (e.g., distributed within a three-dimensional (3D) sphere with minimum inter-node distance restriction); new physical conditions might need to be considered in the analysis and design, e.g., the random jitter of UAV will affect the channel characteristics, and raise challenge for channel state information (CSI) acquisition in practice.

\item{\em New constraints:} Energy efficiency becomes more important as the UAV systems are in general energy-limited; propulsion energy consumption needs to be considered, which is related to the UAV's velocity and 
acceleration rate; practical kinetic models also pose new constraints on the feasible positions when the trajectory of a UAV is to be designed.
\end{itemize}

Concerning of these new issues appeared in the field of UAV-PLS, a number of excellent survey works have been done  \cite{Sun_2019UAV87,Wu_IWC_UAV48,Shang_ICM_UAV64,Li_IWC_UAV13,Wang_IWC_UAV45}. 
In these works, important research directions were summarized, possible application scenarios were described, and future opportunities and challenges were discussed. However, there is still lack of a systematic and comprehensive review of the large amount of existing technical contributions in the literature up to present. 
To this end, we try to establish a clear framework for the study of UAV-PLS, and under this framework, provide a comprehensive review of the most important existing works and latest research progresses. Note that in the existing survey works of UAV communications, e.g., \cite{UAVsurvey2015,Fotouhi_survey19,Mozaffari_survey19,Hayat_survey16}, the PLS might have been mentioned but not as a major concern. To our best knowledge, this is the first comprehensive survey focusing on the PLS issue in UAV communication systems. 

\begin{figure}[t]
	\centering
	\includegraphics[width=0.9\columnwidth]{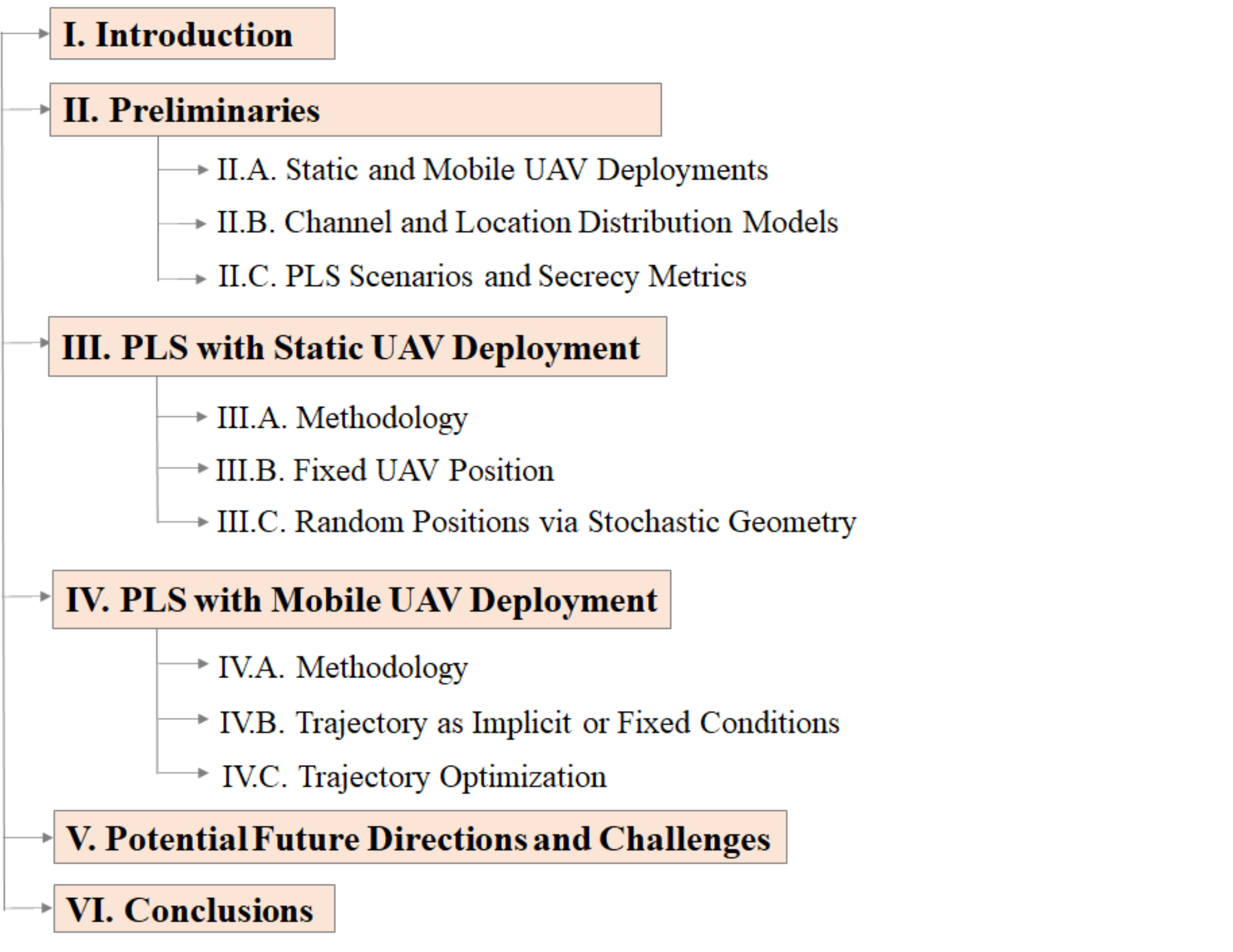}
	\caption{Organization of this survey.}
	\label{overview}
\end{figure}

This survey is structured as Fig.~\ref{overview}. In Section II, we first present necessary preliminaries for the subsequent discussion. Note that many existing works on UAV-PLS are largely fragmented: The UAV may play different roles, the secrecy metrics are various, and the design problem ranges from conventional resource allocation/power control/precoding design, to UAV-specific problems such as position/trajectory optimization. To describe these dispersed works in an organized way, we classify them into two major categories according to the deployment form of UAVs, namely static and mobile deployments. 
These two deployment forms generally have different application scenarios and analysis/design methodologies, which will be explained briefly in {\em Subsection II.A}, and later described in detail in Section III and IV, respectively. In {\em Subsection II.B}, we highlight the unique characteristics of A2G channels, as well as the location distribution models in UAV systems. These specific models render the analysis and design of UAV-PLS  in general different as compared to that in the conventional terrestrial scenarios. In {\em Subsection II.C}, we summarize the possible roles of UAV under the concept of PLS. Specifically, UAV may act as a legitimate communication node (transmitter (Tx) or receiver (Rx)), a helper (friendly jammer or relay), an attacker, or a monitor/hidder in the concept of covert communications. For different roles, the design objectives and secrecy metrics are in general different, which are described therein accordingly.   

In Section III and Section IV, we survey the existing works for static and mobile UAV deployments, respectively. For each section, we first summarize the general design objectives and methodologies commonly adopted for the corresponding deployment form (in {\em Subsection III.A} and {\em Subsection IV.A}), then provide detailed description of the corresponding works in the literature. The surveyed works are further categorized according to the analysis framework or design methodology they used. Specifically, for static UAV deployment, we consider two major research directions: 1) ({\em Subsection III.B}) The UAV and/or the other communication nodes have fixed positions, and hence performance analysis and transmission design are conducted under a fixed typology. 2) ({\em Subsection III.C}) The positions of the communication nodes are randomly distributed in a target area. In this case, the spatially averaged long-term secure performance is the focus. Similarly, the literature related to mobile UAV deployment is also surveyed for two major research categories: 1) ({\em Subsection IV.B}) The UAV flies along a pre-determined fixed trajectory, or along a trajectory not being explicitly specified. For this branch of research direction, the major objective is to investigate the impact of UAV moving on the system performance, while the trajectory itself is not a design objective. 2) ({\em Subsection IV.C}) The trajectory of UAV is to be optimized. The so-called trajectory planning problem becomes a major research direction. It has been investigated for various scenarios and contributes the most number of works to the literature.

In Section V, we further introduce some important emerging directions in the field of UAV-PLS. Some cutting-edge techniques including reconfigurable intelligent surface (RIS), UAV swarm-enabled coordination, integrated UAV-satellite networks, mobile edge computing (MEC), and machine learning, etc., are envisioned to enhance the PLS of UAV communications in the coming sixth generation (6G) era.
At last, concluding remarks are provided in Section VI.

\section{Preliminaries}

\subsection{Static and Mobile UAV Deployments}

\subsubsection{\textbf{Static Deployment}}
In some application scenarios, UAV may be deployed at fixed locations to  provide coverage for a target area, when the terrestrial communication infrastructure is absent, damaged, or its capacity becomes insufficient due to user crowding. When only one ground receiver is considered, the optimal UAV position, obviously, should be placed as closely as possible to the receiver when UAV acts as a transmitter, or it should be determined jointly considering the locations of the transmitter and the receiver when UAV acts as a relay \cite{Chen_CL18, Fan_CL18}. When multiple ground users need to be served, an important design objective is to guarantee coverage with minimum cost, e.g., in terms of the power consumption or the required number of UAVs. This can be done by optimizing the UAV's 3D position $(x_{\rm U},y_{\rm U},z_{\rm U})$ \cite{Hourani-WCL, Alzenad_WCL18}. Furthermore, when multiple UAV transmitters are available, guaranteeing ground coverage can be formulated as a so-called disk covering problem.
As the coverage region of the UAV forms a circular disk on the ground, the problem aims to cover the target area with minimum number of disks with minimum cost, which has been widely investigated for the conventional communication systems in the literature \cite{Alzenad_WCL17, Mozaffari_TWC16,Lyu_CL17}.

\begin{figure}
	\centering 	
		\subfigure[Static deployment.]
{\includegraphics[width = 0.41\textwidth]{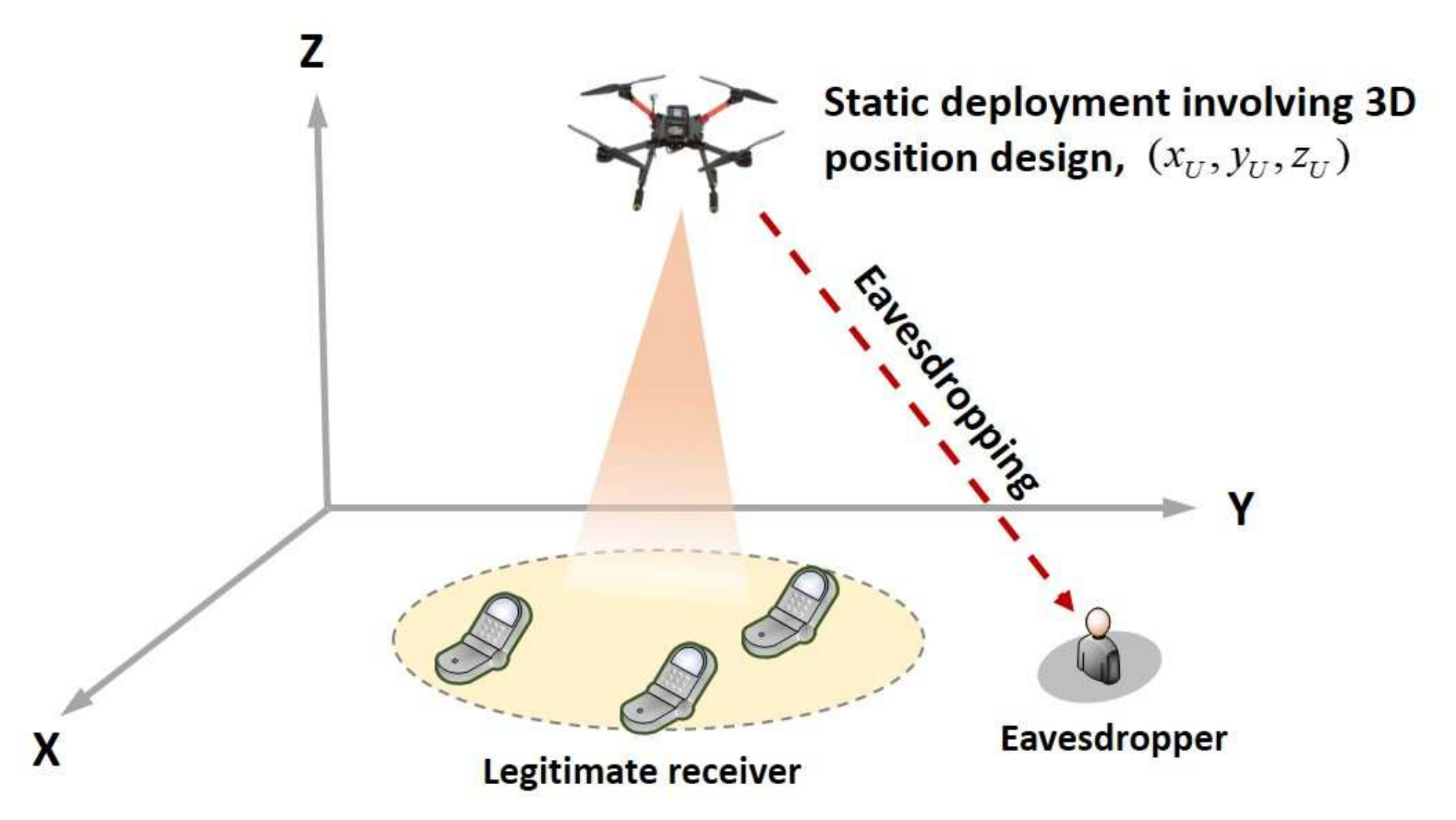}\label{3D_position}}		
	
	\subfigure[Mobile deployment.]
{\includegraphics[width=0.41\textwidth]{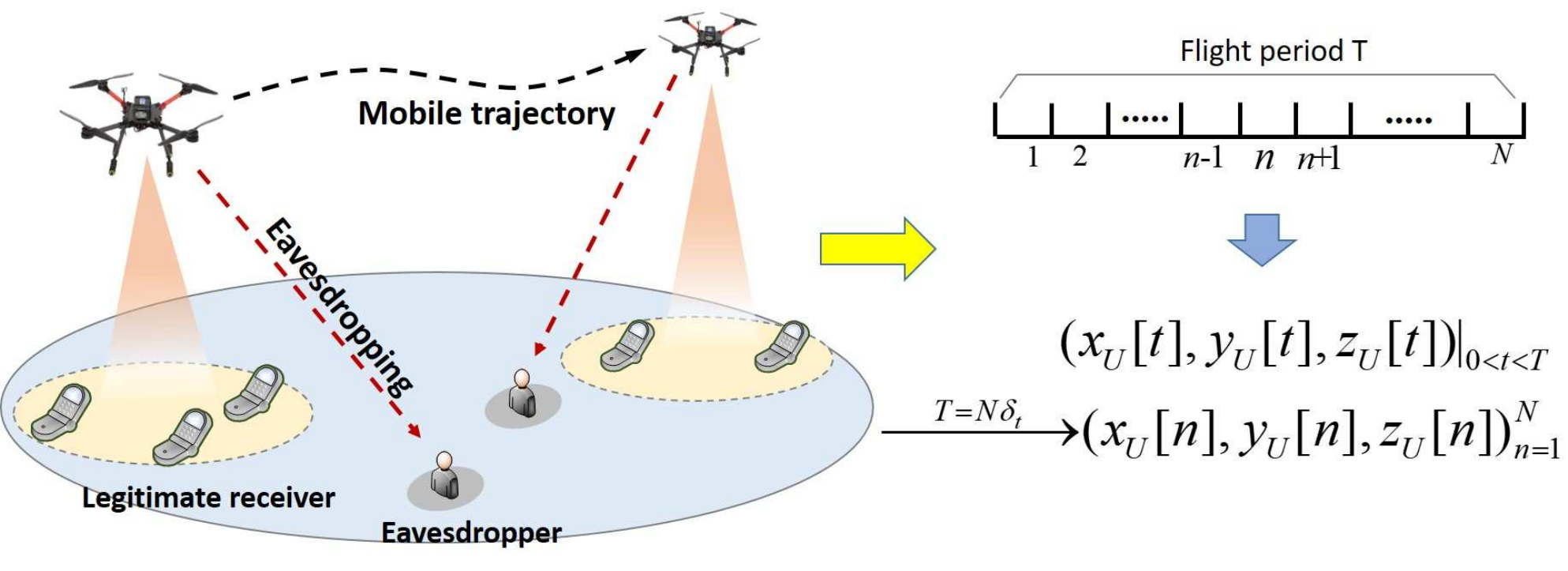}\label{trajectory}}			

			\centering  	
			\caption{Example scenarios of secure UAV communications with static/mobile deployment.}
			\label{unique_characteristic}
\end{figure}

As far as PLS is concerned, the UAV position should be determined jointly considering the locations of both the legitimate nodes and malicious attackers, as illustrated by an example scenario in Fig.~\ref{3D_position}. In general, the UAV should be placed close to the legitimate nodes, and away from the malicious nodes, to guarantee channel superiority for the legitimate links to realize information-theoretical secure communication. However, the problem usually cannot be so easily formulated and solved. On the one hand, the location information of passive eavesdroppers (Eves) cannot be easily obtained. Although it has been assumed that the position of Eves can be detected by the UAV, via an optical camera or synthetic aperture radar (SAR) equipped on-board, see e.g., \cite{Zhang2019-UAV50}, the acquired Eve's location could be imperfect. Besides, the fast fading channel of Eve, if considered, in general cannot be known for passive eavesdropping. In this case, designing secure communication without (or with partial of) Eve's CSI is practically important. 
On the other hand, possible trade-off, which is not that straightforward, may exist if some sophisticated channel models are considered. For example, when blocking probability in the A2G channel is taken into account, increasing the UAV altitude $z_{\rm U}$ may reduce the blocking probability, meanwhile, the path loss is also increased. These effects need to be jointly considered for both the legitimate and malicious links, based on that, the UAV altitude should be carefully designed according to the secrecy performance metric of interest. Moreover, when multiple ground nodes (either legitimate or malicious) are considered, the aforementioned disk-covering problem will become more complex. These design issues will be reviewed and discussed in detail in the following Section III.

It should be emphasized, that for more practical consideration, a UAV in general cannot be perfectly statically deployed. UAV hovering may cause shadowing effect \cite{Bao_TVT_UAV161}, and inevitable UAV jitter may introduce new randomness to the channel \cite{Wu2020-UAV141}. These effects will be specifically highlighted when we introduce the A2G channels in Subsection II.B.

\subsubsection{\textbf{Mobile Deployment}}

Unlike the rotate-wing UAV which can stay at a fixed position during the service time, a fix-wing UAV has to keep flying. For example, it may fly along a circular route over the target area \cite{Chen_CL18}, or it may fly from an initial location to a final location (take-off and landing sites) according to predefined flight route \cite{WangWCL2017-UAV135}. For these cases, the trajectory of UAV is considered as a fixed condition, other than an optimization dimension. When the flight route is fixed and known a priori, it is possible to exploit this information for efficient transmission design. As an example, the large-scale fading conditions over all time slots along the route now can be predicted. With such information, the transmission can be designed in a long-term manner, e.g., allocating power over all time slots during the flight via water filling-like approaches.

The trajectory of UAV could also be designed on purpose. For example, the user to be served is moving and the UAV is deployed for accompanying flight, or there exist multiple users separated in distance, such that the UAV needs to fly from one to another to accomplish data transmission/collection with completion time and quality of service (QoS) constraints. When PLS is considered, the flight trajectory should be designed avoiding the Eves' locations. Different from static UAV deployment, now a series of positions over a continuous time duration need to be optimized. 
A commonly adopted analysis framework is transmission time 
discretization. As shown in Fig.~\ref{trajectory}, the entire flight period $T$ is 
divided into $N$ equal-length time slots, where $T=N\delta_t$ with $\delta_t$ representing an unit time slot. It is usually assumed that $\delta_t$ is sufficiently small such that the UAV can be approximately considered to be static within each slot. Via discretization, the UAVs' continuous trajectory $(x_{\rm U}[t],y_{\rm U}[t],z_{\rm U}[t])$, $0<t<T$, can be approximately expressed as $(x_{\rm U}[n],y_{\rm U}[n],z_{\rm U}[n])_{n=1}^N$, where $(x_{\rm U}[n],y_{\rm U}[n],z_{\rm U}[n])$ denotes the UAV's Cartesian coordinate in the $n$-th time slot. In practice, the length of $\delta_t$ should be properly chosen, accounting for the trade-off between approximation accuracy and implementation complexity.

When optimizing $(x_{\rm U}[n],y_{\rm U}[n],z_{\rm U}[n])_{n=1}^N$, different positions in different slots will be correlated according to UAV kinetic constraints. Usually the constraints rely on the initial location, the final location, and the maximum speed of the UAV, $v_{\rm max}$. As an example where the UAV is assumed to fly at a fixed altitude (i.e., $z_{\rm U}[n] = H, \forall n$) such that only the 2D trajectory is designed,\footnote{When 3D trajectory is considered, i.e., $z_{\rm U}[n]$ also needs to be optimized, the complexity of the problem will be further increased \cite{Sun_TCom19}.} these constraints are expressed as \cite{Zhang2019-UAV50}
\begin{multline}\label{eq:1}
(x_{\rm U}[n+1] - x_{\rm U}(n))^2 + (y_{\rm U}[n+1] - y_{\rm U}[n])^2 \leq (v_{\rm max} \delta_t)^2, \\
n = 1, ..., N - 1,
\end{multline}
where $(x_{\rm U}[0], y_{\rm U}[0])$ and $(x_{\rm U}[N], y_{\rm U}[N])$ denote the initial and final coordinates of the UAV, respectively, for which the following constraint should be applied to guarantee the feasibility of the problem.
\begin{equation}\label{eq:2}
(x_{\rm U}[N] - x_{\rm U}(0))^2 + (y_{\rm U}[N] - y_{\rm U}[0])^2 \leq (v_{\rm max} T)^2.
\end{equation}
Another common constraint on the initial and final position is to guarantee that the UAV return to its take-off place. In this case, the constraint reads as \cite{Cai_JASC_UAV9}
\begin{equation}\label{eq:3}
(x_{\rm U}[N], y_{\rm U}[N]) = (x_{\rm U}[0], y_{\rm U}[0]).
\end{equation}

There exist other analysis and design frameworks. In \cite{Zhou_ICSPCS17}, the UAV is moving but transmits only at fixed positions, termed as ``stop points''. Given a target area and a flight route, the number and locations of the stop points are to be optimized. In some other works, UAV moving is implicitly reflected in the system model without being directly related to its physical positions. For example, in \cite{Liu_ICUFN17}, UAV mobility will affect the corresponding link reliability, which is mathematically modelled as a Bernoulli process. These approaches will be discussed in detail later in Subsection IV.

\subsection{Channel and Location Distribution Models}

For both static and mobile UAV deployments, it is in general challenging to acquire instantaneous CSI of all involved nodes. Some reasons are as follows: The number of users (both legitimate and malicious) in the vast coverage area of UAV would be large; for the mobile deployment case, the fast moving of UAV results in highly dynamic fading channel, and will require very frequent CSI update. Fortunately, the A2G channel (and the A2A channel as well) has been recognized to be sparse scattering. In this case, the large-scale path loss dominates the channel gain, as has been widely assumed in the literature. As long as the location of a node is known, large-scale fading can be evaluated given a certain path loss model,  and the PLS design can be conducted accordingly. In this subsection, we introduce the commonly adopted channel model and location distribution models for UAV-PLS. For a more comprehensive survey of the existing UAV channel modelling approaches, the readers are referred to \cite{Khuwaja_survey18,Khawaja_UAV203,Yan_Access2019}.

\subsubsection{\textbf{Path Loss Models}}
The A2G channels exhibit unique characteristics as compared to conventional terrestrial propagation channels. This may cause differences in the secrecy performance analysis and secure communication design.
In general, the A2G channels behaves differently in different scenarios, e.g., urban, rural, or over-water areas \cite{Yan_Access2019}, for which various models have been proposed in the literature \cite{Yun_Access2015,Goddemeier_GCWorkshop2010, Ibrahim_milcome2015,Gulfam_ICC2015}. In most of the existing models, LoS propagation plays an important role, which is very likely to happen when UAVs are deployed at high altitudes. Nevertheless, the LoS link could be blocked in practice, and the blockage probability would be related to the UAV altitude. In general, higher altitude corresponds to larger LoS probability. Considering this effect, a probabilistic LoS/NLoS A2G channel model has been widely considered, e.g., in \cite{Hourani-WCL,Liu_TWC_UAV88,Liu_UAV2}, and described as 
 \begin{equation}
 \label{PL_avg}
 [\text{PL}_{\text{Avg}}]_{\rm dB}=\mathcal{P}_{\rm SU}^{\text{LoS}}\times[\text{PL}_{\text{LoS}}]_{\rm dB}+\mathcal{P}_{\rm SU}^{\text{NLoS}}\times[\text{PL}_{\text{NLoS}}]_{\rm dB},
 \end{equation} 
where $\mathcal{P}_{\rm SU}^{\text{LoS}}$ denotes the LoS probability between a ground node (${\rm S}$) and UAV (${\rm U}$), which is given by
 \begin{equation}
 \label{P_LoS}
 \mathcal{P}_{\rm SU}^{\text{LoS}}=\frac{1}{1+a \exp\big[-b(\theta_{\rm SU} -a)\big]},
 \end{equation}
 where $a$ and $b$ are constants relying on the propagation environments, and $\theta_{\rm SU}$ is the elevation angle, expressed as
 \begin{equation}
 \theta_{\rm SU} = \frac{180}{\pi}\arcsin\frac{z_{\rm U}}{d_{\rm SU}}.
 \end{equation}
Correspondingly, the NLoS probability is  
$\mathcal{P}_{\rm SU}^{\text{NLoS}}=1-\mathcal{P}_{\rm SU}^{\text{LoS}}$.

$[\text{PL}_{\text{LoS}}]_{\rm dB}$ and $[\text{PL}_{\text{NLoS}}]_{\rm dB}$ in \eqref{PL_avg} are the path loss components for the LoS and NLoS conditions, respectively. These two terms are determined in different ways in the literature. In some works, they are simply assumed to be free space path loss (FSPL) with different offsets. Written in a standard log-distance form, these are given by
\begin{align}
  [\text{PL}_{\text{LoS}}]_{\rm dB}&=20 \log d_{\rm SU}+20 \log \frac{4 \pi f_c}{c}+\eta_{\text{LoS}},\label{eq:4}\\
   [\text{PL}_{\text{NLoS}}]_{\rm dB}&=20 \log d_{\rm SU}+20 \log \frac{4 \pi f_c}{c}+\eta_{\text{NLoS}}, \label{eq:5}
 \end{align}
where
   $f_c$ is the carrier frequency, $c$ is the speed of light, $\eta_{\text{NLoS}}$ and $\eta_{\text{LoS}}$ denote the constant offsets to the LoS/NLoS FSPL, respectively. In other works, the LoS and NLoS path loss components may have different path loss exponents, not only being different in the offsets \cite{Chen_Access_UAV41,Sun2019-UAV39}. For a more sophisticated modelling approach, the path loss parameters could be elevation angle-dependent \cite{Tatar_UAV154}. Specifically, the path loss exponent may decrease with the elevation angle. To describe this effect, it was suggested that the path loss exponent can be modelled as a decreasing function of $\theta_{\rm SU}$. An example can be found in \cite{Azari_TCom18}.

Note that the non-linear sigmoid form of LoS probability in \eqref{P_LoS} poses challenges in the analysis and performance evaluation. To tackle this problem, a piecewise-fitting model was proposed in \cite{Tang_TIFS_UAV105}, where the LoS probability was described as a piecewise linear function of the propagation distance between a UAV transmitter and a ground receiver. With this model, further analysis can be effectively facilitated.

The LoS/NLoS condition of the A2G channel link has been described in alternative ways. In \cite{Yao2019-UAV29}, it is assumed that when the elevation angle is larger than a threshold, i.e., the user is located inside a disk area below the UAV, then LoS propagation dominates the channel. In the urban scenario, the LoS probability can also be determined according to the location and height distribution of the buildings \cite{Kang_CL2019,Holis_TAP08}. Specifically, the locations of buildings are modelled as Poisson point process (PPP) on the ground, with their heights following Rayleigh distribution. In such an environment, whether the LoS link between a UAV and a ground user is blocked can be then evaluated accordingly.

Finally, as for ${\rm PL}_{\rm LoS}$ and ${\rm PL}_{\rm NLoS}$ in \eqref{PL_avg}, there exist more complex and specific A2G models in the literature, e.g., the floating intercept path loss model, the dual slop path loss model, the two-way path loss model, and the modified FSPL model, etc. \cite{Khawaja_UAV203}. In practice, these models should be carefully selected according to the scenarios of interest.

\subsubsection{\textbf{Shadowing and Small-Scale Fading}}

Although it has been widely assumed that the large-scale path loss dominates the performance in UAV systems, shadowing and small-scale fading need to be considered for some specific scenarios. Similar to the path loss parameters, it has been shown that the shadowing might also be affected by the elevation angle of the channel link \cite{Hourani_GC14}.
Besides the external environment, recent measurements have shown that the airframe itself could also cause severe shadowing which cannot be neglected. During aircraft maneuvering, the shadowing effect may last for a duration up to dozens of seconds, which has to be taken into account in some mission-critical operations \cite{Sun_ICNSC}.

For small-scale fading, most conventional stochastic models, including Rayleigh, Rician, Nakagami-$m$, Weibull fading models, etc., have been considered in the A2G scenario \cite{Cui_Access2020}. Among them, Rician fading has been widely assumed in the presence of a dominant LoS link. Again, the $K$-factor could be elevation angle-dependent \cite{Azari_TCom18}. Rician fading has also been considered for modelling the A2A channel link in 
\cite{Goddemeier_GC15}.

\subsubsection{\textbf{Location Distribution Models}}
When explicit CSI is not known, the location distribution information of both the legitimate and malicious nodes is important in the secure communication design, especially for the case when the large-scale path loss dominates the system performance. In the literature, PPP is usually adopted for describing the distribution of the ground nodes -- this basically has no obvious difference compared to that has been commonly assumed for terrestrial communication systems. However, the location distribution of UAVs shows some difference. When modelling the location distribution of a UAV swarm, it is necessary to consider an extra constraint that the inter-UAV distance should be larger than a safety threshold. To address this constraint, some specific point process, e.g., the Mat\'{e}rn Hardcore point process, has been introduced in the literature \cite{Zhu_JSAC_UAV40}. Besides, different from the ground nodes, the UAVs can be deployed in a 3D space. In \cite{Yapici_UAV152}, it was assumed that the position of UAV is uniformly distributed in a horizontal distance range, and also uniformly distributed in a altitude range.

Being different to the stochastic geometry-based location distribution models, some works consider practical UAV mobility when describing its random location. For example, in \cite{Bhattacharya_10}, a kinematic model has been applied to describe the location evolution of UAV. In \cite{Sharma_TWC19}, random waypoint mobility (RWPM) and uniform mobility (UM) models were adopted to describe the UAV location as a 3D-mobility movement process. The readers are referred to these papers and the references therein for more information on this research direction. 

At last, we introduce some widely adopted location error models, which are usually considered for robust design in the cases that the location of Eve cannot be precisely obtained. A bounded location error model has been considered in \cite{Zhou_TC_UAV158}, where the real location coordinate $\mathbf{q}$, is modelled by the known location coordinate $\bar{\mathbf{q}}$ plus an error vector $\Delta\mathbf{q}$, such that
\begin{equation}
\mathbf{q} = \bar{\mathbf{q}} + \Delta\mathbf{q},
\end{equation}
where $\Delta\mathbf{q}$ could be randomly drawn from an uncertainty region $\Psi$, defined as
\begin{equation}
\Psi \triangleq \{\Delta\mathbf{q} \in \mathbb{R}^{2 \times 1}: ||\Delta\mathbf{q}||^2 \leq \omega^2\},
\end{equation}
where $\omega$ is the radius of the error disk. For the ease of analysis, $\omega$ was modelled as a Gaussian random variable with zero mean and non-zero variance in \cite{Zhou_TSP19}. More information on the location error model is referred to the references therein.

 \subsection{PLS Scenarios and Secrecy Performance Metrics}
As the last but an important part of the preliminary, we summarize the commonly considered PLS scenarios and the corresponding secrecy performance metrics. According to the role that a UAV may act, we classify the PLS scenarios into four categories: 1) UAV acts as communication node (transmitter or receiver), 2) UAV acts as helper (friendly jammer or relay), 3) UAV acts as attacker (malicious jammer or Eve), and 4) UAV acts as hider or surveillant (in the concept of covert communications). While some secrecy performance metrics are common for all these scenarios, there exist specific performance metrics which are more concerned in a certain scenario, as will be discussed and highlighted in the following.

\subsubsection{\textbf{UAV as Communication Node}}
As communication nodes, the UAV may act as either legitimate transmitter or receiver. In the presence of eavesdropping, a major design objective is to maximize the secrecy rate (SR), which has been widely considered in the study of PLS and defined as
\begin{equation}
\label{SecRate_Rs}
   R_s=[R_m-R_e]^+,
\end{equation}
where $R_m=\text{log}_2(1+\gamma_m)$ and $R_e=\text{log}_2(1+\gamma_e)$ are the achievable rate of the main channel and the wiretap channel, respectively, with $\gamma_{m}$ and $\gamma_{e}$ respectively representing the instantaneous SNRs of the legitimate link and the eavesdropping link. As long as $R_s > 0$, wiretap coding can be conducted to guarantee zero information leakage to Eve. 

However, evaluating \eqref{SecRate_Rs} requires knowledge of Eve's instantaneous CSI, which is usually unavailable. Fortunately, in UAV communications, the channel and system performance are usually dominated by the large-scale fading due to sparse scattering, as previously discussed. In this case, \eqref{SecRate_Rs} can be calculated with location information only, and optimal design to maximize the SR can be conducted. Specifically for the mobile UAV deployment where UAV trajectory is to be designed, average secrecy rate (ASR) is usually considered as the optimization objective.\footnote{The ASR should not be confused with the ergodic secrecy rate (ESR), where the latter is obtained by taking expectation over all possible fading states of the involved channel links. Besides, $\frac{1}{N}$ is usually omitted when evaluating the ASR when $N$ is fixed as a constant. In this case, the ASR is equivalent to the accumulate SR.} The system designer aims to maximize $\sum_{n = 1}^{N} R_s[n]$, where $R_s[n]$ is the SR in the $n$-th discrete time slot and $N$ denotes the total number of slots during the considered flight period. A general form of the optimization problem is described as
\begin{align}
{\rm maximize}&~~~~ \sum_{n = 1}^N R_s[n],\label{eq:12}\\
{\rm s.t.}&~~~~ \mathbf{q}_{\rm U}[n], \mathbf{P}[n], {\rm ~and ~other ~constraints.}
\end{align}
where $\mathbf{q}_{\rm U}[n] = (x_{\rm U}[n], y_{\rm U}[n], z_{\rm U}[n])$ denotes the location coordinate of UAV in slot $n$, usually constrained by the previously described \eqref{eq:1}--\eqref{eq:3}, $\mathbf{P}[n]$ is the power allocation at slot $n$. There may exist other practical constraints according to the problem of concern, e.g., constraints on the UAV kinematics such as acceleration.

In addition to the SR, energy consumption is an important issue that needs to be specifically addressed in UAV communications, since the on-board energy budget is usually limited. For this consideration, secrecy energy efficiency (SEE) has been considered. It is defined as the ratio between the amount of secret bits that are successfully transmitted and the total energy consumption used to support such transmission, 
        \begin{equation}\label{eq:SEE}
        	\eta_{\rm SEE}=\frac{R_s}{E_{\rm total}} ~(\text{bits/J}),
        \end{equation}
where $E_{\rm total}$ represents the total energy consumed by the UAV, which may include not only the communication energy, but also the energy consumed for hovering. For the moving UAV deployment, being different from the conventional terrestrial communication scenario, it is assumed that the communication energy consumption is much less than the propulsion energy of UAV. In this case, $E_{\rm total}$ in \eqref{eq:SEE} solely depends on the UAV mobility parameters such as velocity and acceleration, while it is irrelative with the communication power consumption $\mathbf{P}[n]$ \cite{Hua_TVT2019,Xiao-TGCN-UAV44}. However, in the case that the trajectory is pre-fixed and cannot be controlled by the communication provider, the propulsion energy is excluded from the total energy \cite{Bai_TVT_UAV69}.

When explicit CSI is not available but its distribution information is known, secrecy outage probability (SOP) is another commonly adopted performance metric, which measures the probability that the SR drops below a pre-determined threshold. In UAV communications, SOP is considered when the system designer knows the locations of all nodes but cannot know their instantaneous fading, or when the designer cannot know the explicit locations but only knows the location distribution.
Denote SOP as $\mathcal{P}_{\rm out}$ and it can be expressed as 
   \begin{equation}\label{eq:SOP}
   \mathcal{P}_{{\rm out}}=\mathbb{P}\{R_s<R_{th}\},
   \end{equation}
 where $R_s$ is the achievable SR defined in \eqref{SecRate_Rs}, and $R_{th}$ is the target SR threshold.
 
Some variant forms of the outage probability has been discussed, e.g., the hybrid outage probability (HOP) which jointly considers the outage occurred in the main link and the outage induced by eavesdropping \cite{Liu_TWC_UAV88}, the intercept probability (IP) which describes the probability that the SNR of the legitimate link is less than that of the eavesdropping link \cite{Bao_TVT_UAV161}, and the secrecy connection probability (SCP) which describes the probability of non-zero secrecy capacity \cite{Tang_TIFS_UAV105}. In practice, outage analysis is usually followed by artificial noise (AN)-aided jamming design, which has been shown to be a practical solution for secure transmission when explicit CSI is unknown. As previously discussed, this constitutes an important research direction in the field of UAV-PLS, as explicit CSI acquisition in UAV communications is always challenging.

\subsubsection{\textbf{UAV as Helper}}
The UAV may act as helper to aid the secure communication between other communication nodes. When acting as a relay, the design objective basically has no much difference -- the SR \eqref{SecRate_Rs}, ASR \eqref{eq:12}, SEE \eqref{eq:SEE}, and SOP \eqref{eq:SOP} are still widely used as performance metrics in the analysis and design.

When UAV acts as a friendly jammer, new performance metrics appear for practical scenarios that the location information of Eve is unknown. In this case, the designer may try to reduce the outage probability in a suspicious target area. The union of all locations where jamming can be beneficial, that is, jamming helps to reduce the outage probability, is defined as jamming coverage \cite{Vilela_TIFS11}. In practice, the design objective could be maximizing the jamming coverage area \cite{Zhou_ICSPCS17}.
Similarly, an intercept security area was defined as the union of locations where Eve's signal to noise and interference ratio (SINR) is smaller than a threshold. Maximizing this area has been considered in \cite{Zhou_TVT_UAV17}. For a friendly jammer who cannot know the information of Eve's location, these area-based performance metrics provide practical and effective design objectives.

\subsubsection{\textbf{UAV as Attacker}}
It is important to consider malicious UAVs in the study of UAV-PLS. Flexible deployment and mobility of malicious UAV will render the attacking more powerful and harmful. Besides, the attacker may benefit from the high probability of LoS propagation. These should be specifically taken into account in the secure transmission design.  
In most cases, the previously described secure performance metrics are still interested. Differently, now the impacts of new parameters such as the altitude of the UAV attacker, are specifically emphasized in the analysis. For example, the secrecy performance has been compared for two cases (Eve is on the ground or in the air) in \cite{Abughalwa_ICTC19}, and the impact of UAV height on the SOP has been investigated in \cite{Ye_GC18}. 

New constraints, objectives, or design framework also arise in this scenario. For example, when a UAV transmits directionally to cover a disk area on the ground, the high-risk region for eavesdropping can be described as a cone below the transmitter. To guarantee PLS, 3D constraint on Eve's location should be considered for the case of UAV Eve. More specifically, the design should avoid any Eve appearing in this cone region below the legitimate transmitter \cite{WuICCC-UAV145}. Moreover, considering a smart UAV attacker, its power, position, or moving trajectory could be dynamically adjusted to maximize its attacking efficiency. In this case, game-theoretical approaches are usually applied in the analysis and secure transmission design, see e.g., \cite{Xu_Access18, Li_Access18, Li_Eurosip20}.

\subsubsection{\textbf{Covert Communications and Wireless Surveillance}}
As an important brunch in the studies of PLS, covert communications aim to hide legitimate transmission from detecting, while surveillance aims to identify the existing of unauthorized or malicious transmitters. The use of UAV brings opportunities and challenges for both: High probability of LoS channel links is beneficial for legitimate surveillance, while it becomes more challenging for a transmitter to hide itself. 

Hypothesis testing is usually conducted for this scenario. With a certain detection rate being guaranteed, the false alarm rate and miss detection rate are commonly adopted performance metrics. These are respectively defined as follows,
\begin{align}
\mathbb{P}_{F} &= {\rm Pr}\{\mathcal{D}_1|\mathcal{H}_0\},\\
\mathbb{P}_{M} &= {\rm Pr}\{\mathcal{D}_0|\mathcal{H}_1\},
\end{align} 
where $\mathcal{D}$ and $\mathcal{H}$ denote the decision and hypothesis, respectively, with the subscripts $0$ (or $1$) representing there does not exist (or exist) an unauthorized transmission. A common design objective is to determine an optimal detection threshold, which minimizes the overall detection error rate. When UAV is involved, the position or trajectory of UAV can be further optimized for this purpose \cite{Zhou_TSP19,Hu_PC_21}.

\section{PLS with Static UAV Deployment}

In this section, we review the existing literature on PLS with static UAV deployment. We first introduce the commonly adopted methodologies in this area, then describe the related works under two distinct research frameworks, respectively. For the first category, UAV will be deployed at fixed positions, which might be further optimized. For the second category, the positions of UAVs (and/or other nodes) are considered to be random following certain spatial stochastic process, e.g., PPP. In this case, the long-term performance evaluated over the possible location distribution of the involved nodes, as well as the impact of key distribution parameters (such as the node density) on PLS, are of particular interest.

\subsection{Methodology}

\subsubsection{\textbf{Common Analysis and Optimization Framework}}

As previously discussed, static UAV deployment has been widely considered for scenarios such as coverage enhancing for the cell edge and temporary hotspot areas. Basically, this scenario does not have much difference as compared to the conventional terrestrial communications, except for the following aspects:
\begin{itemize}

\item
Unique A2G/A2A channels lead to new performance analysis results. A notable observation is that the performance may highly rely on the elevation angle of the propagation link, which may affect the LoS probability, the PL exponent, the Rician $K$-factor, etc., and subsequently result in different secrecy performance.

\item
UAVs can be flexibly deployed in 3D space. This introduces a new dimension that can be exploited in the design of PLS. The 3D position optimization problem is in general non-convex, however, via proper transformation and approximation techniques, the original problem usually can be re-formulated as solvable problems. For example, in \cite{Wang_WCL19,Wang_UAV167}, the 3D position optimization problem was formulated as quadratically constrained quadratic fractional
programming (QCQFP), which was transformed into quadratically
constrained quadratic programming (QCQP), and further approximated then solved with semi-definite relaxation (SDR).

\item
A UAV attacker could be more harmful. Specifically, in addition to conventional actions that a smart attacker can do, such as dynamic power adjustment and attacking strategy selection, a UAV attacker can further dynamically change its position for more effective attacking. For this scenario, game theory is usually applied to analyze the secrecy performance under smart attacks. For a survey of the game theoretic approaches in UAV systems, the readers are referred to \cite{Mkiramweni_survey19}.
\end{itemize}

\subsubsection{\textbf{Stochastic Geometry}}
The other research direction focuses on randomly distributed nodes, and aim to evaluate the average secrecy performance for a random network. For this direction, conventional stochastic geometry theory provides a powerful tool. 
With the widely-adopted PPP model, the number of nodes within an interested area, as well as the inter-node distances, can be conveniently described by their distribution functions. Accordingly, the PLS performance can be evaluated in a stochastic manner, e.g., in terms of outage probability, or SR averaged over spatially random locations. For modelling the UAVs' positions, the distribution model might be further extended to 3D. Besides, new models are applied to account for the minimum inter-UAV distance requirement, such as the MHC point process.

\subsection{Fixed UAV Position}

For clear exposition, we summarize the PLS works with fixed UAV positions according to the roles of UAV, as has been described previously in Subsection II.C. 

\subsubsection{\textbf{UAV Transmitter/Receiver}}
We start from the works where UAV acts as transmitter (i.e., UAV-mounted BS) or receiver. In \cite{Wang_WCL19}, a basic three-node system model was considered, where a UAV BS transmits to a ground legitimate receiver (Bob), in the presence of a ground Eve. It was assumed that the position of all nodes are known, whereas the instantaneous fading channels are not. Therefore, non-zero secrecy capacity probability, i.e., the SCP described in Subsection II.C, was considered as the performance metric. The 3D position of UAV BS was optimized with a two-step approach, which first optimizes the UAV altitude and then the horizontal position. The same design objective and optimization problem has been considered in \cite{Wang_UAV167}, while with an extra practical constraint of obstacle avoidance. Multiple obstacles on the ground were modelled as semiellipsoids, and the deployment position of UAV should maintain a minimum distance to these obstacles in the optimization. 
UAV receiver was considered in \cite{Liu_TWC_UAV88,Liu_UAV2}, where a multi-antenna ground source transmits to a single-antenna UAV, in the presence of a multi-antenna ground Eve. Jamming is performed at both the source and full-duplex (FD) Eve. HOP was analyzed, and the optimal configurations of both the source and Eve were discussed according to the UAV altitude.

When multiple antennas are implemented at the UAV, spatial dimension can be exploited and there will be more to be designed aside of the UAV position. For example, spatial multiplexing techniques can be adopted to serve more users on the same time-frequency resource, beamforming can be utilized to concentrate the signal power to a target user while reducing information leakage, and AN can be sent in the null space of legitimate channel links to degrade the Eve's link quality. In \cite{Wang_TCom2020_Tang}, a multi-antenna UAV BS was used to serve multiple ground users with non-orthogonal multiple access (NOMA) and simultaneous wireless information and power transfer (SWIPT). Thanks to the antenna array implemented at the UAV, AN can be generated to jam the ground Eve simultaneously with legitimate information transmission. The sum rate of all users are maximized to find the optimal precoding and AN jamming vectors, as well as the SWIFT parameters such as the power splitting ratio. 

For a multi-antenna UAV, UAV jitter becomes an important issue. Even when the position of the communication nodes can be known, the random jitter, which cannot be precisely predicted, may cause uncertainty in the knowledge of the LoS channel direction. Specifically, jitter will cause random variations in the azimuth and elevation angles of the LoS link, which will subsequently degrade the system performance. Fortunately, the maximum angle variation range can be guaranteed. According to this information, a robust design takes the worst-case performance into consideration. For example, the most harmful jitter angle is assumed for the legitimate node, and the most beneficial jitter angle is assumed for attackers.  Together, these assumptions were used to evaluated the worst-case secrecy performance, and further adopted as design objective or performance requirement in \cite{Wu2020-UAV141,Wen_WCNC20}.

\subsubsection{\textbf{UAV Relay/Jammer}}
The UAV may work as helper to aid the secure communications between legitimate nodes. i.e., as relay or friendly jammer. In this case, the position of UAV and the relaying/jamming strategy should be jointly considered. For UAV relaying, a basic four-node model was discussed in \cite{Bao_TVT_UAV161}, where a decode-and-forward (DF) UAV relay is deployed to aid the communication between a ground BS and Bob, in the presence of a ground Eve. Analytical expressions of the intercept probability and ESR were derived, and accordingly, the impact of the UAV altitude/power on the secrecy performance was discussed via numerical simulations.

Considering that the UAV helper is energy-constrained, efforts have been devoted to the area of SWIPT to allow the UAV harvest energy from information or jamming signals.
In \cite{Tatar_UAV154}, a UAV relay receives energy from information-bearing signal sent by the ground source, as well as jamming signal sent by a FD destination. 
A more complex scenario has been considered in \cite{Wang_TCom20}, where  the UAV relay and legitimate receiver are both FD devices, thereby they can transmit and simultaneously perform jamming against multiple Eves on the ground. Robust design was conducted considering uncertainty in the Eves' location information, and the position of UAV, together with jamming and SWIPT protocols, are jointly optimized to maximize the worst-case SR.

More effective relaying can be realized with multiple antennas implemented at the UAV, or when there exist multiple UAVs cooperatively aiding the transmission. With a multi-antenna UAV relay, a common design objective is to jointly optimize the information and AN precoders \cite{LiBin-2018-UAV21}, similar to the previous case that UAV acts as a multi-antenna BS \cite{Wang_TCom2020_Tang}. When multiple UAV helpers are involved in the transmission, secrecy problem may arise from the inside -- that is, UAV could be untrustworthy. The consideration is practical, since it is in general challenging to monitor the behaviour of every individual UAV in a swarm. For such a scenario, AF protocol should be applied, and the design involves intended jamming and relay selection, with which SOP and ESR have been analyzed in \cite{Nuradha_Arxiv19}. In \cite{Hosseinalipour_20}, the problem has been investigated from the attacker's perspective. Considering a multi-hop transmission assisted by multiple UAV relays, the optimal jammer's location, which is most harmful to the legitimate transmission, was analyzed. The work has opened up a new thinking direction -- knowing the location which is most vulnerable to attacking can help the system designer make better prevention.

Aside of acting as relay, the UAV can be deployed as friendly jammer to improve the secrecy performance of legitimate communications. Mobility of the UAV can be exploited for efficient jamming. Clearly, if the location of Eve is known, it is better to place the UAV jammer as close as possible to Eve. Although it has been widely assumed that the Eve's location can be detected by UAV via, e.g., optical camera onboard \cite{Caris_AESMag14}, considering unknown/imperfect location information of Eve is of practical importance. In \cite{Zhou_ICSPCS17}, a basic four-node system model, i.e., including one source, one destination, one Eve and one UAV jammer, was considered. Without knowing Eve's location, the jamming coverage was defined as the area within which jamming can be beneficial to reduce the SOP, and the design aims to maximize the jamming coverage area by optimizing the UAV jammer's position. A similar system model was considered in \cite{Zhou_TVT_UAV17}, where also an area-based metric, namely the intercept probability security region (IPSR) which describes the union of all positive security regions, has been used as the design objective to be maximized. The discussion on UAV friendly jamming was further extended to the classic multiple-input single-output multiple-Eve (MISOME) scenario in \cite{Chen_Access_UAV41}. With multiple antennas implemented at the friendly jammer, null space beamforming can be conducted to avoid the jamming interfering with legitimate transmission links. Jamming power allocation and UAV position optimization have been investigated to maximize the transmission rate, constrained on a target SOP requirement.

\subsubsection{\textbf{UAV Attacker/Suspicious UAV}}
As compared to ground attackers, a UAV attacker benefits from both flexible and agile deployment, as well as better channel conditions. In \cite{Abughalwa_ICTC19}, flying and ground Eves have been compared, trying to overhear the communication between a UAV Tx and a ground Bob. Considering various fading characteristics of the A2A/A2G channels, SOP and ESR have been investigated. Accordingly, it was shown that flying Eve in general could be more harmful. 

When the UAV attacker is smarter, i.e., it can dynamically adjust its attacking strategy, game-theoretical approaches are usually applied.
In \cite{Li_Access18}, the UAV attacker may select from four action modes, namely eavesdropping, jamming, spoofing, or keeping silent, according to the transmission strategy adopted by Alice. On the other hand, to combat with the UAV attacker, Alice dynamically determines its transmission power. Via game-theoretical analysis, the ASR as well as the attacking rates, including the eavesdropping rate, the jamming rate, and the spoofing rate, have been investigated through simulations.
Game-theoretical analysis has been applied in a UAV network in \cite{Xu_ICCC19}, where the Tx/Rx and attacker (jammer) are all UAVs. Taking into account the mutual-interference in multi-pair UAV communications, the legitimate UAV Tx's and the UAV jammer dynamically adjust their transmit power to maximize their utilities, respectively, where the process is formed as a Stackelberg game. Also for a UAV network, cooperative relaying has been investigated in \cite{Xu_TVT20_game}, where the user association was formulated as a many-to-one matching game, and cooperative transmission was formulated as an overlapping coalition formation game, considering both selfish users and malicious UAV Eves.

In some cases, UAV is regarded as suspicious communication node, and the design is conducted from the eavesdropper's (legitimate monitor) perspective. Related works are termed as legitimate eavesdropping \cite{Hu_CL20} or
proactive eavesdropping \cite{Lu_TVT19,Wu_EURASIP19}. In \cite{Hu_CL20}, the legitimate monitor aims to wiretap on the communication between a pair of suspicious Tx and Rx, which is aided by a suspicious UAV relay. The monitor performs jamming to guarantee that it has higher receive SNR than the suspicious Rx, therefore the suspicious messages are decodable at the monitor. However, the UAV relay can dynamically adjust its position to alleviate the impact of jamming, this has been taken into account when determining the optimal jamming power at the monitor. A similar scenario has been investigated in \cite{Lu_TVT19}, wherein the suspicious communication is between a UAV Tx and multiple ground receivers. In this case, the UAV Tx dynamically adjusts its position given a certain jamming environment, to maximize the minimum achievable rate among all receivers, and the monitor accordingly determines its optimal jamming power to maximize the eavesdropping rate.

\subsubsection{\textbf{Covert Communications and Wireless Surveillance}}
Covert communications and wireless surveillance have been widely investigated for UAV systems. In the concept of covert communications, UAV aims to hide its transmission from a malicious detector; while for wireless surveillance, UAV is adopted to monitor (or conduct legitimate eavesdropping on) the suspicious communications occurred in the environment.

For covert UAV communications, the LoS/NLoS condition of the detecting link plays an important role in the analysis.
A basic three-node system model has been considered in \cite{Yan_ICC19}, where a UAV Tx wants to hide its transmission from a ground warden (Willie), which locates right below the UAV. The probabilistic LoS/NLoS model is elevation angle-dependent, hence the link from UAV to Willie is always LoS, while the transmission link to the legitimate receiver, Bob, has LoS or NLoS conditions depending on the UAV altitude. Hypothesis testing is conducted at Willie, and the UAV Tx optimizes its altitude and power, aiming to maximize the SNR at Bob while guaranteeing the detection probability under a certain threshold. The work has been further extended in \cite{Yan_Arxiv20} where 2D UAV position was optimized. Investigation has been conducted from Willie's perspective in \cite{Hu_Wu_TVT20}, where a more powerful Willie implemented with multiple antennas tries to identify the existence of UAV communications using beam sweeping. The optimal number of sectors for beam sweeping was discussed based on closed-form expressions of the detection error probability. 

For UAV-aided wireless surveillance, existing works are mainly conducted from the perspective of legitimate eavesdropping. That is, the UAV monitor aims to decode the suspicious message, instead of only detecting the existence of suspicious communications. A basic three-node model was considered in \cite{Shen_ICCWksp20}, where a UAV monitor wants to decode the message sent by a suspicious Tx, in the meanwhile, it performs jamming to the suspicious Rx. This is similar to the proactive eavesdropping works described previously; however, with UAV being the legitimate monitor, its position can be flexibly adjusted to maximize the surveillance performance \cite{Shen_ICCWksp20}. Analysis of UAV surveillance has been extended to a four-node scenario \cite{Hu_Access19}, where the suspicious communication between source and destination is aided by a suspicious relay. It is also considered for a UAV network in \cite{Li_TVT19_Sur}, where the legitimate monitor, the suspicious communication nodes are all UAVs. From an opposite perspective, techniques against UAV surveillance have been discussed in \cite{Wang_TCom20_covert}.

For the ease of reader's reference, we summarize some representative PLS works with fixed UAV positions in Table~\ref{tab:addlabe0},\footnote{Different terminologies might be adopted in different works. For transmitter, it might be denoted as BS, Tx, source, Alice, etc. For receiver, it could be denoted as Rx, destination, Bob, etc. For convenience, in Table~\ref{tab:addlabe0}, we keep consist with the term used in the corresponding original paper.} where the concerned scenarios, the roles of UAV and malicious nodes, as well as their position assumptions, the channel characteristics, the secrecy-related metrics, and the design objectives are highlighted, respectively. This may help the reader quickly understand the latest progress in his/her interested research direction.

\begin{table*}[htbp]
\tiny
  \caption{UAV-PLS works with fixed UAV position.}
  \centering
    \begin{threeparttable}
    \footnotesize
  \resizebox{\textwidth}{!}{
    \begin{tabular}{|c|c|c|c|c|c|c|c|c|}
    \hline
    \multicolumn{1}{|c|}{\multirow{2}[4]{*}{Ref.}} & \multicolumn{1}{c|}{\multirow{2}[4]{*}{Scenario}} & \multicolumn{2}{c|}{UAV} & \multicolumn{2}{c|}{Malicious nodes} & \multicolumn{1}{c|}{\multirow{2}[4]{*}{Channel characteristics}} & \multicolumn{1}{c|}{\multirow{2}[4]{*}{Metric}} & \multicolumn{1}{c|}{\multirow{2}[4]{*}{Objective}} \\
\cline{3-6}          &       & \multicolumn{1}{c|}{Role} & \multicolumn{1}{c|}{Position} & \multicolumn{1}{c|}{Role} & \multicolumn{1}{c|}{Position} &       &       &  \\
    \hline
\cite{Wang_WCL19}      &  \tabincell{c}{{Air: UAV BS (1)\tnote{1}}\\{Ground: Bob (1), Eve (1)}}     &   BS    &   To be optimized    &   Eve    &  \tabincell{c}{{Fixed}\\{Known}}	  &   \tabincell{c}{{NLoS}\\{Rayleigh fading}}    &  SCP & Optimize UAV position\\
    \hline
\cite{Wang_UAV167} & \tabincell{c}{{--\tnote{2}}\\{(\textbf{Ground obstacles})\tnote{3}}} & -- & -- & -- & -- & -- & -- & --\\
\hline  
\cite{Liu_TWC_UAV88}& \tabincell{c}{{Air: UAV Rx (1)}\\{Ground: BS (1), Eve (1)}}& Rx & Fixed & \tabincell{c}{{Eve}\\{(with jamming)}} & \tabincell{c}{{Fixed}\\{Known}}& \tabincell{c}{{Probabilistic LoS/NLoS}\\{Rayleigh (NLoS), Rician (LoS)}}& HOP & \tabincell{c}{{Performance analysis}}\\
\hline
\cite{Wang_TCom2020_Tang} & \tabincell{c}{{Air: UAV BS (1)}\\{Ground: Bob ($K$), Eve (1)}\\{({\bf Multi-antenna UAV,}}\\{{\bf SWIPT, NOMA})}}& -- & Fixed & Eve & \tabincell{c}{{Fixed}\\{Unknown}} & -- & Throughput	 & \tabincell{c}{{Optimize precoding,AN vectors,}\\{and SWIPT protocol}}\\
\hline 
\cite{Wu2020-UAV141} &\tabincell{c}{{Air: UAV BS (1)}\\{Ground: Bob (1), Eve (1)}\\{({\bf UAV jitter})}}&--&\tabincell{c}{{Fixed}\\{(with jitter)}}&--&\tabincell{c}{{Fixed}\\{Known}}&\tabincell{c}{{Rician (LoS + NLoS)}\\{LoS component: steering vector}}&\tabincell{c}{{Worst-case SR}\\{(as constraint)}}& \tabincell{c}{{Optimize beamforming,}\\{and AN vectors}}\\
\hline
\cite{Wen_WCNC20} &--&--&--&--&--&--&\tabincell{c}{{Worst-case SR}\\{(as objective)}}& --\\
\hline
\cite{Bao_TVT_UAV161} &\tabincell{c}{{Air: UAV relay (1)}\\{Ground: BS (1), Bob (1),}\\{and Eve (1)}}& DF Relay& \tabincell{c}{{Hovering}\\{(cause shadowing)}}&--&--&\tabincell{c}{{FSPL}\\{Log-normal shadowing}\\{Fading ignored}}&\tabincell{c}{{Inception Prob.}\\{ESR}}& Performance analysis\\
\hline
\cite{Tatar_UAV154}&\tabincell{c}{{--}\\{({\bf Cooperative jamming,}}\\{{\bf SWIPT})}}&\tabincell{c}{{AF relay}\\{(with SWIPT)}}&Fixed&--&--&\tabincell{c}{{Probabilistic LoS/NLoS}\\{Rician}\\{(angle-dependent $K$-factor)}}&\tabincell{c}{{Connection Prob.}\\{SOP}\\{ESR}}&Performance analysis\\
\hline
\cite{Wang_TCom20} &\tabincell{c}{{Air: UAV relay (1)}\\{Ground: BS (1), Bob (1),}\\{and Eve ($n$)}\\{({\bf Cooperative jamming,}}\\{{\bf SWIPT})}}&--&To be optimized&--&\tabincell{c}{{Fixed}\\{Known}\\{(with error)}}&\tabincell{c}{{ FSPL}\\{Fading ignored}}&Worst-case SR&\tabincell{c}{{Optimize UAV position, }\\{SWIPT, and jamming}}\\
\hline
\cite{LiBin-2018-UAV21} &--&AF relay&Fixed&--&\tabincell{c}{{Fixed}\\{Known}}&\tabincell{c}{{Probabilistic LoS/NLoS}\\{Rayleigh (NLoS), Rician (LoS)}}&--\tnote{4}& \tabincell{c}{{Optimize beamforming,}\\{ and AN vectors}}\\
\hline
\cite{Nuradha_Arxiv19} &\tabincell{c}{{Air: UAV relay ($n$)}\\{Ground: Source (1),}\\{and Destination (1)}\\{({\bf Untrustworthy UAV})}}& -- & -- & \tabincell{c}{{Untrustworthy}\\{ relay}\\{(UAV)}} &\tabincell{c}{{Fixed}\\{Known}}&\tabincell{c}{{Probabilistic LoS/NLoS}\\{Nakagami-$m$ fading}}&SOP, ESR& \tabincell{c}{{Jamming design,}\\{ and relay selection;}\\{Performance analysis}} \\
\hline
\cite{Hosseinalipour_20} &--&DF relay &--& Jammer& To be optimized& \tabincell{c}{{LoS (A2A), NLoS (G2G, G2A);}\\{Fading ignored}}& SIR & Optimize jammer's location\\
\hline
\cite{Zhou_ICSPCS17} &\tabincell{c}{{Air: UAV jammer (1)}\\{Ground: BS (1), Bob (1),}\\{and Eve (1)}}&Friendly jammer&To be optimized &Eve&Unknown & \tabincell{c}{{Probabilistic LoS/NLoS}\\{Fading ignored}}&\tabincell{c}{{SOP}\\{Jamming coverage}}&\tabincell{c}{{Optimize UAV's position}}\\
\hline
\cite{Zhou_TVT_UAV17} &--&--&--&--&--&--&IPSR&\tabincell{c}{{Optimize UAV's position,}\\{ and power}}\\
\hline
\cite{Chen_Access_UAV41} &\tabincell{c}{{Air: UAV jammer (1)}\\{Ground: BS (1), Bob (1),}\\{and Eve ($M$)}\\{({\bf MISOME})}}&--&--&--&\tabincell{c}{{Fixed}\\{Known}}&\tabincell{c}{{Probabilistic LoS/NLoS}\\{Rayleigh (NLoS), Rician (LoS)}}&\tabincell{c}{{SOP}\\{(as constraint)}}& \tabincell{c}{{Optimize UAV's position,}\\{ and jamming power}}\\
\hline
\cite{Abughalwa_ICTC19} &\tabincell{c}{{Air: UAV Tx (1),}\\{and UAV Eve (1)}\\{Ground: Bob (1), Eve (1)}}& \tabincell{c}{{Tx}\\{(and Eve)}} & Fixed &\tabincell{c}{{Eve}\\{(Ground/UAV)}}& -- & --&SOP, ESR & Performance analysis\\
\hline
\cite{Li_Access18} &\tabincell{c}{{Air: UAV attacker (1)}\\{Ground: Alice (1), Bob (1)}\\{({\bf Game theory})}}& \tabincell{c}{{Attacker}\\{(multi-mode)}} & N.A. & \tabincell{c}{{Multi-mode}\\{ attacker}\\{(UAV)}} &N.A.&\tabincell{c}{{Imperfect estimation of}\\{instantaneous channel}}&\tabincell{c}{{ASR}\\{Attacking rate}}&\tabincell{c}{{Mode selection for Attacker;}\\{Power adjustment for Alice}\\{(via game-theoretical analysis)}}\\
\hline 
\cite{Xu_ICCC19} &\tabincell{c}{{Air: UAV Tx/Rx}\\{ (multi-pair)}\\{UAV jammer (1)}\\{({\bf Game theory})}}&\tabincell{c}{{Tx/Rx}\\{Jammer}}&N.A.& \tabincell{c}{{Jammer}\\{(UAV)}} & N.A.& Perfectly known&N.A.&\tabincell{c}{{Power adjustment for UAV Tx,}\\{and for UAV jammer}\\{(via Stackleburg game)}}\\
\hline
\cite{Xu_TVT20_game} &\tabincell{c}{{Air: UAV Tx ($N$), Rx ($M$)}\\{UAV Eve ($S$)}\\{({\bf Game theory})}}&\tabincell{c}{{Tx/Rx}\\{Eve}}&Fixed&\tabincell{c}{{Eve}\\{(UAV)}}&\tabincell{c}{{Fixed}\\{Known}}& \tabincell{c}{{FSPL}\\{Fading ignored}}&Worst-case SR&\tabincell{c}{{User association,}\\{Cooperative transmission}}\\
\hline
\cite{Hu_CL20} &\tabincell{c}{{Air: UAV (1)}\\{Ground: Tx (1), Rx (1),}\\{and Monitor (1)}\\{({\bf Legitimate eavesdropping})}}&\tabincell{c}{{Relay}\\{(suspicious)}} &\tabincell{c}{{Adjustable}\\{Known}}& \tabincell{c}{{Eve/Jammer}\\{(legitimate)}}& Fixed &--& \tabincell{c}{{Eavesdropping}\\{ rate}}& \tabincell{c}{{Optimize jamming power}}\\
\hline
\cite{Lu_TVT19} & \tabincell{c}{{Air: UAV (1)}\\{Ground: Rx ($K$), Monitor (1)}\\{({\bf Legitimate eavesdropping})}}& \tabincell{c}{{Tx}\\{(suspicious)}} &--&--&--&--&--&--\\
\hline
\cite{Yan_ICC19,Yan_Arxiv20} &\tabincell{c}{{Air: UAV Tx (1)}\\{Ground: Bob (1), Willie (1)}\\{({\bf Covert communications})}}& Tx &To be optimized & Willie & -- &\tabincell{c}{{Probabilistic LoS/NLoS}\\{AWGN}}&\tabincell{c}{{Detection prob.}\\{(as constraint)}}&\tabincell{c}{{Optimize UAV's position,}\\{ and power}}\\
\hline
\cite{Hu_Wu_TVT20} &\tabincell{c}{{Air: UAV Tx (1)}\\{Ground: Willie (1)}\\{({\bf Covert communications})}\\{({\bf Multi-antenna Willie})}}& Tx & Fixed &--& Fixed &-- &\tabincell{c}{{Detection error}\\{prob.}}& \tabincell{c}{{Optimize sweeping sectors}\\{(of Willie)}}\\
\hline
\cite{Shen_ICCWksp20} &\tabincell{c}{{Air: UAV (1)}\\{Ground: Tx (1), Rx (1)}\\{(suspicious)}\\{({\bf UAV surveillance})}}&\tabincell{c}{{Legitimate}\\{monitor}}&Adjustable & \tabincell{c}{{Tx/Rx}\\{(suspicious)}} &\tabincell{c}{{Fixed}\\{Known}}&\tabincell{c}{{Probabilistic LoS/NLoS}\\{Rician}\\{(angle-dependent $K$-factor)}}&\tabincell{c}{{Surveilling}\\{non-outage prob.}}& Performance analysis\\
\hline
\cite{Hu_Access19} &\tabincell{c}{{Air: UAV (1)}\\{Ground: Suspicious Tx (1),}\\{relay (1), and Rx (1)}\\{({\bf UAV surveillance})}}& -- &To be optimized &\tabincell{c}{{Tx/Rx, relay}\\{(suspicious)}} &--& Rayleigh &\tabincell{c}{{Average}\\{surveilling rate}}& \tabincell{c}{{Optimize UAV's position,}\\{ and jamming}}\\
\hline
    \end{tabular}}
\begin{tablenotes}
  \item[1] Number in the bracket indicates the number of this kind of nodes.
  \item[2] For brevity, we use ``--'' to denote ``Same as above'' in the tables.
  \item[3] Special scenarios will be emphasized with bold font in bracket.
  \item[4] ``worst-case SR'' may refer to different meanings in different works. It could be measured over uncertainty of jitter/Eve location, or over multiple Eves.
\end{tablenotes}
\end{threeparttable}
  \label{tab:addlabe0}%
\end{table*}%

\subsection{Random Positions via Stochastic Geometry}

Stochastic geometry provides a powerful tool for performance analysis in random networks. In this case, the system performance is evaluated in a spatially large-scale manner, i.e., averaged over space. As far as UAV-PLS is concerned, stochastic point processes have been assumed for Eves, legitimate communication nodes, or UAVs in a swarm. We review the related approaches in this subsection.
According to the different analysis frameworks adopted, we structure this subsection into two parts. Namely, 1) UAV has fixed position while the other nodes are randomly located, and 2) multiple UAVs in a swarm are randomly distributed in space. Similar to previous discussion, different roles of UAV have been assumed. This will be emphasized when necessary.

\subsubsection{\textbf{Random Network with Fixed UAV Position}}
For this research direction, UAV is deployed in a random network, where the positions of legitimate communication nodes and/or attackers are randomly distributed. The position of UAV, however, is considered to be fixed or to be optimized, according to stochastic geometry-based secrecy performance analysis. 

When UAV acts as legitimate Tx, the classic three-node wiretap channel model has been extended to multiple-Eve scenario in \cite{Omri_VTC18}, where the Eves are randomly located on ground following PPP. SOP was analyzed and the impacts of system parameters such as the UAV altitude and Eves' density have been discussed. In addition to ground Eves, multiple UAV Eves were considered in \cite{Ye_GC18}. While the ground Eves are randomly distributed within a circle on ground, UAV Eves are randomly distributed within a sphere in 3D space. SOP has been investigated for this scenario. Multiple Bobs, multiple Eves, and multiple buildings in an urban scenario have been considered in \cite{Kang_CL2019}. The locations of ground buildings are also modelled by PPP with their height following Rayleigh distribution. With this building-blockage model, LoS probability was evaluated, the average worst-case SR (AWSR) was analyzed, and the UAV altitude has been optimized accordingly. A multi-antenna UAV BS was considered in \cite{Rapasinghe_Asilomar18}, which performs 3D beamforming to serve randomly distributed ground users with NOMA, in the presence of randomly distributed ground Eves. A novel idea of ``protected zone'' was proposed, which is defined as an area around the user that has been cleared from Eves. Aiming to maximize the sum SR, the optimal shape of the protected zone has been discussed.

Analysis has been extended for different roles of UAV, e.g., receiver \cite{Lei_ICCC19, Lei_IOT20, Mobini_ICC18}, relay \cite{Sun2019-UAV39,Sun_WCL_UAV31}, or jammer \cite{Tang_TIFS_UAV105}. Among these research works, the general methodologies adopted are similar, i.e., describing the inter-node distance distribution according to their location distribution, and further relating the distance to secrecy performance metrics. However, there may exist slight difference. For example, the UAV altitude would affect the stochastic properties of both legitimate and eavesdropping links when UAV is a transmitter, while it only affects the legitimate link when UAV acts as receiver. 

\begin{table*}[tbp]
\tiny
  \caption{UAV-PLS analysis via stochastic geometry.}
  \centering
    \begin{threeparttable}
    \footnotesize
  \resizebox{\textwidth}{!}{
    \begin{tabular}{|c|c|c|c|c|c|c|c|c|}
    \hline
    \multicolumn{1}{|c|}{\multirow{2}[4]{*}{Ref.}} & \multicolumn{1}{c|}{\multirow{2}[4]{*}{Scenario}} & \multicolumn{2}{c|}{UAV} & \multicolumn{2}{c|}{Malicious nodes} & \multicolumn{1}{c|}{\multirow{2}[4]{*}{Channel characteristics}} & \multicolumn{1}{c|}{\multirow{2}[4]{*}{Metric}} & \multicolumn{1}{c|}{\multirow{2}[4]{*}{Objective}} \\
\cline{3-6}          &       & \multicolumn{1}{c|}{Role} & \multicolumn{1}{c|}{\tabincell{c}{{Position or}\\{Distribution}}} & \multicolumn{1}{c|}{Role} & \multicolumn{1}{c|}{\tabincell{c}{{Position or}\\{Distribution}}} &       &       &  \\
    \hline
\cite{Omri_VTC18} &\tabincell{c}{{Air: UAV (1)}\\{Ground: Bob (1), Eves (M)}}& Alice & Fixed & Eve & PPP & \tabincell{c}{{Probabilistic LoS/NLoS}\\{Log-normal shadowing}\\{Rayleigh (NLoS), Rician (LoS)}}& SOP & Performance analysis\\
    \hline
\cite{Ye_GC18} &\tabincell{c}{{Air: UAV Tx (1), UAV Eves (M)}\\{Ground: Rx (1), Eves (M)}}& Alice/Eve &\tabincell{c}{{Alice: Fixed}\\{Eve: PPP}}& \tabincell{c}{{Eve}\\{(Ground/UAV)}} & -- & \tabincell{c}{{A2A: LoS}\\{A2G: LoS/NLoS}\\{Rayleigh (NLoS), no fading (LoS)}}&--&--\\
\hline 
\cite{Kang_CL2019} &\tabincell{c}{{Air: UAV (1)}\\{Ground: Bobs (M), Eves (M)}\\{({\bf Building blockage})}}&Alice & \tabincell{c}{{Altitude}\\{to be optimized}}& Eve & -- & \tabincell{c}{{Probabilistic LoS/NLos}\\{(via building blockage model)}\\{Fading ignored}}&AWSR & \tabincell{c}{{Optimize UAV altitude}}\\
\hline
\cite{Rapasinghe_Asilomar18} &\tabincell{c}{{--}\\{({\bf Protected zone,}}\\{{\bf multi-antenna UAV, NOMA})}}& -- & Fixed & -- & -- &\tabincell{c}{{LoS channel: steering vector}\\{FSPL}\\{Fading ignored}}& Sum SR & Optimize protected zone \\
\hline
\cite{Lei_ICCC19} &\tabincell{c}{{Air: UAV (1)}\\{Ground: Alice (1), Eves (M)}}& Bob & -- & -- & -- & \tabincell{c}{{Probabilistic LoS/NLoS;}\\{Rician}\\{(angle-dependent $K$-factor)}}&SOP/ASR & Performance analysis\\
\hline
\cite{Lei_IOT20} &\tabincell{c}{{Air: UAV (2)}\\{Ground: Alice (1), Eves (M)}\\{({\bf Dual UAV})}}  & Bob/Jammer & -- & -- & -- & -- & -- & -- \\
\hline
\cite{Mobini_ICC18} &\tabincell{c}{{Air: UAV (1)}\\{Ground: Suspicious Tx (1),}\\{Rx (M), and Monitor (1)}\\{({\bf Proactive eavesdropping})}}& Rx & -- &\tabincell{c}{{Suspicious}\\{nodes}}& -- &\tabincell{c}{{G2A: LoS, fading ignored}\\{G2G: Rayleigh}}& \tabincell{c}{{Non-outage prob.}\\{at Monitor}}& \tabincell{c}{{Optimize UAV altitude,}\\{Monitor power,}\\{ and beamformer}}\\
\hline
\cite{Sun2019-UAV39} &\tabincell{c}{{Air: UAV (1)}\\{Ground: Source (1), Destination (1),}\\{ and Eves (M)}\\{({\bf SWIPT, mmWave})}}& AF/DF Relay & -- & Eve & -- & \tabincell{c}{{G2A/A2G: LoS}\\{G2G: mmWave blockage model}\\{Fading: Gamma distributed}\\{(power gain)}}&\tabincell{c}{{ASR}\\{Energy coverage}}& Performance analysis\\
\hline
\cite{Sun_WCL_UAV31} & -- & DF Relay & To be optimized & \tabincell{c}{{--}\\{(Colluding)}} & -- & -- & ASR & \tabincell{c}{{Optimize transmit power,}\\{UAV position,}\\{and power splitting ratio}}\\
\hline
\cite{Tang_TIFS_UAV105} &\tabincell{c}{{Air: UAV Jammer (1), Eves (M)}\\{Ground: Tx (1), Rx (1)}}& Jammer/Eve & \tabincell{c}{{Jammer: Fixed}\\{Eve: BPP}}& \tabincell{c}{{--}\\{(UAV)}} & BPP & \tabincell{c}{{A2A: LoS, FSPL}\\{A2G: Probabilistic LoS/NLoS}\\{A2A/A2G: Fading ignored}\\{G2G: Rayleigh}}& SCP & Performance analysis \\
\hline
\cite{Shi_Phycom19} &\tabincell{c}{{Air: UAVs ($K$)}\\{Ground: Source (1), Users ($M$),}\\{and Eves ($L$)}\\{({\bf Caching})}}& \tabincell{c}{{DF relay}\\{(with cache)}}& PPP & \tabincell{c}{{--}\\{(Colluding)}} & PPP & \tabincell{c}{{Normal PL model}\\{Rayleigh fading}}&\tabincell{c}{{Secure}\\{cache}\\{throughput}} & Optimize caching prob.\\
\hline
\cite{Ma_access19} & \tabincell{c}{{Air: UAVs (M)}\\{Ground: Source (1), Destination (1),}\\{and Eves (M)}\\{({\bf mmWave})}}&\tabincell{c}{{DF relay}\\{Friendly Jammer}}& \tabincell{c}{{PPP}\\{(Fixed height)}}& -- & -- &\tabincell{c}{{Probabilistic LoS/NLoS}\\{Nakagami-$m$ fading}}&SOP & Performance analysis\\
\hline
\cite{Yao2019-UAV29} &\tabincell{c}{{Air: UAVs (M)}\\{Ground: Bobs (M), Eves (M)}}& BS & -- & -- &-- &\tabincell{c}{{LoS/NLoS}\\{(elevation angle-dependent)}\\{LoS: FSPL, no fading}\\{NLoS: Large PL exponent (4),}\\{Rayleigh fading}}&\tabincell{c}{{SOP}\\{Secrecy capacity}}& \tabincell{c}{{Optimize wiretap coding rate,}\\{UAV altitude, guard zone}}\\
\hline
\cite{Zhu_JSAC_UAV40} &\tabincell{c}{{Air: UAVs (M)}\\{Ground: Bob (1), Eves (M)}\\{({\bf MHC process})}}& \tabincell{c}{{BS}\\{Friendly jammer}} & \tabincell{c}{{MHC process}\\{(Fixed height)}}& -- & -- & \tabincell{c}{{Probabilistic LoS/NLoS}\\{Nakagami-$m$ fading}\\{3D antenna gain}}& \tabincell{c}{{CDF of SINR}\\{ASR}} & Performance analysis\\
\hline
\cite{Yapici_UAV152} & \tabincell{c}{{Air: UAV ($K$)}\\{Ground: BS (1)}\\{({\bf 3D distribution, protected zone})}}& Rx/Eve & \tabincell{c}{{Uniform distribution}\\{(in horizontal distance,}\\{and height)}}&\tabincell{c}{{--}\\{(UAV)}}& Uniform distribution & \tabincell{c}{{Probabilistic LoS/NLoS}\\{Rician (LoS), Rayleigh (NLoS)}}& SR & --\\
\hline
\cite{Ye_WCL_UAV36} & \tabincell{c}{{Air: UAV Tx (1), Rx (1)}\\{UAV Eves ($N$)}\\{({\bf UAV network, 3D distribution})}}& Tx/Rx/Eve & PPP & \tabincell{c}{{--}\\{(UAV)}}& PPP &\tabincell{c}{{LoS, FSPL}\\{No fading}}& SOP/ASC & --\\
\hline
\cite{Bao2020wcl} & \tabincell{c}{{Air: UAV Rx (1)}\\{UAV Eves ($M$)}\\{Ground: BS (1)} \\ {(\textbf{Multi-antenna BS})}}& Rx/Eve & \tabincell{c}{{Rx: Fixed} \\ {Eve: PPP}} & \tabincell{c}{{--}\\{(UAV, colluding)}}& PPP &\tabincell{c}{{LoS (A2G} \\ {$k-\mu$ fading)}}& SOP & {Performance analysis}\\
\hline

    \end{tabular}}
\end{threeparttable}
  \label{tab:addlabel}%
\end{table*}%

\subsubsection{\textbf{Random UAV Positions}}
When multiple randomly located UAVs are concerned, the modelling of their position distribution shows some difference as compared to the modelling for ground nodes. First, UAVs may be distributed in 3D space; moreover, to avoid UAV collision, minimum distance between UAVs needs to be guaranteed for practical consideration.

As a commonly adopted simplified scenario, all UAVs can be assumed to distribute within a disk area at certain height. In this case, 2D point processes still can be applied. In \cite{Tang_TIFS_UAV105}, binomial point process (BPP) has been assumed to model $N$ randomly distributed UAV Eves within a disk above the ground Tx. PPP has been used to model the location distribution of multiple UAV relays/jammers \cite{Shi_Phycom19,Ma_access19}, in which the number of UAVs in an interested area is random rather than being fixed. An alternative method is to relate the UAV positions to the positions of ground nodes. In \cite{Yao2019-UAV29}, multiple ground Rx's are randomly distributed following PPP, each associated with a UAV Tx located right above its position. 
Further taking into account the constraint on inter-UAV distance, MHC point process was used to model the distribution of UAV BSs in \cite{Zhu_JSAC_UAV40}, which can be viewed as a dependent thinning process of a standard PPP, with the thinning probability determined by the minimum distance requirement. Therefore, the average number of nodes in an interested area, as well as the inter-node distance distribution can be conveniently deduced using standard methodologies in stochastic geometry, and the secrecy performance can be correspondingly analyzed.

For 3D UAV distribution, different approaches have been used to describe the random UAV positions. In \cite{Yapici_UAV152}, the 3D position distribution of multiple UAV receivers was described in two dimensions: The horizontal distance to the ground BS and the UAV altitude. Both was assumed to be uniformly distributed within a certain range. This approach, albeit being simple and straightforward, does not provide tractable and concise analytical results. Therefore, secrecy performance was discussed in \cite{Yapici_UAV152} mainly based on simulations. Extending PPP to 3D, the secrecy performance of an all-UAV network has been investigated in \cite{Ye_WCL_UAV36}. It was assumed that multiple UAV Eves are randomly distributed within a 3D sphere area centered at the UAV transmitter. The number of Eves within the critical range as well as their distances to the transmitter can be conveniently described, and subsequently, SOP was analyzed for the considered network.

At the end of this subsection, it should be highlighted that in the works where random node distribution is considered, defining an area, where Eves cannot be located in, is another important and practical research direction. Such an area was termed as protected zone in \cite{Rapasinghe_Asilomar18}, and termed as secrecy guard zone in \cite{Yapici_UAV152,Yao2019-UAV29}. Details of these stochastic geometry-related works are summarized in Table~\ref{tab:addlabel}.\footnote{In the table, when the number of nodes is denoted by ``M'' (which stands for ``Multiple''), it indicates no explicit number is assumed, and the node positions are generated from an unbounded point process.}

\subsection{Summary of This Section}

We summarize what we have learnt by now. 
\begin{itemize}
\item There are two major categories for the PLS works with static UAV deployment: The communication nodes have fixed positions, or they are randomly distributed. For both cases, it in general has similar research methodologies as that have been adopted for terrestrial communications, e.g., SR/SOP analysis and optimization techniques, or stochastic geometry-based analysis frameworks. Nevertheless, involving UAV nodes in the network would lead to difference for both.

\item A common and important difference may stem from the unique characteristics of the A2G channel, which is usually considered to be elevation-angle dependent. The elevation angle of a propagation link not only affects its LoS probability, but also has impact on the fading parameters such as the path loss exponent, the Rician $K$-factor, etc. These will subsequently affect the PLS performance when UAV is involved in the communications, as either legitimate or malicious node. 
Besides, the UAV itself may have direct impact on the propagation channel. For example, UAV hovering may cause shadowing to the channel, and UAV jitter may result in uncertainty in the channel conditions. These effects should be taken into account for robust PLS communication design.

\item For the works focusing on fixed node positions, 3D UAV position optimization is of particular interest. The corresponding problem is usually non-convex, however, it has been shown that the problem can be effectively solved with standard transformation and relaxing techniques.

\item Considering random node position distribution, conventional stochastic geometry-based analysis framework can be applied. As a reasonable simplification, it has been widely assumed that multiple UAVs may be distributed at the same height, therefore 2D PPP can be used to model their position distribution. This model can be extended to 3D, and extended to more practical scenarios incorporating the inter-UAV distance constraint via, e.g., the MHC point process.

\item The analysis framework may be different when UAV plays different roles. As far as PLS is concerned, the key point is: Which link (legitimate link or attacking link) will be affected by the A2G channel, or which link has UAV involved and is therefore adjustable. Specifically, when UAV acts as a malicious attacker, it may smartly adjust its position for most efficient attacking. For this case, game theoretical approaches can be applied.

\end{itemize}

\section{PLS with Mobile UAV Deployment}

In this section, we focus on the second important research direction, where the UAVs are moving to provide on-demand service to multiple spatially distinct users (or moving users), other than being statically deployed. Related works are also classified into two major categories. For the first category, the UAV trajectory is not a design objective. It is either implicitly reflected in the analysis, or pre-determined as fixed conditions. The second category involves a most important research direction in UAV communications, i.e., trajectory planning. 
Following a similar structure of the previous section, we first introduce the common methodologies adopted for this research area. After that, existing literature will be reviewed in detail for these two research categories, respectively.

\subsection{Methodology}

\subsubsection{\textbf{Communication Process-Oriented Optimization}}
Different from the static case, UAV communications with mobile deployment often form a temporal  communication process corresponding to the flying period of UAVs, and hence, one has to account for the whole process for optimization. For ease of characterization, this communication process is usually discretized into a series of states.
As the flying duration of UAVs is usually larger than the coherent time of A2G channels, proper CSI prediction methods are necessary for such process-oriented optimization. Fortunately, the large-scale channel fading dominates a typical A2G channel, which is location-dependent and can be predicted for a given trajectory of UAVs. In this case, different channel models and CSI prediction error would lead to different optimization results. As for the performance metric, conventional secrecy metrics can still be adopted, while they now need to be averaged over all the communication states during the UAV  flight. New process-oriented performance metric can also be derived.    

\subsubsection{\textbf{Trajectory Optimization under Kinematics and Energy Constraints}}
For the process-oriented design, trajectory optimization would significantly affect the secrecy performance. Intuitively, the trajectories of UAVs directly affects the channel conditions of both the legitimate and eavesdropping links, rendering it an efficient new dimension for PLS enhancement. In practice, the trajectory is a function of continuous time, which leads to infinite variables and makes the optimization problem intractable. Most works divide the fly duration into many discrete time slots, and optimize the UAV position at each time slot rather than the continuous trajectory (as has been explained in Section II.A).  The trajectory optimization problem is usually non-convex, some optimization methods such as the block coordinate descent (BCD) and successive convex approximation (SCA) methods are often adopted.

In practice, UAV kinematics usually restricts the trajectory design. Particularly, both the velocity and acceleration cannot be arbitrarily set, whereas these parameters have their practical limits, respectively. These constraints will make the UAV positions in adjacent time slots correlated, complicating the optimization. 
Another practical constraint is the total energy of UAVs for both flying and communications. Considering the energy efficiency issue, the trajectory that is optimal for PLS enhancement might requires more propulsion energy consumption. Possible trade-off should be exploited to strike the balance between performance and energy consumption.

\subsection{Trajectory as Implicit or Fixed Conditions}

\subsubsection{\textbf{Implicit Trajectory}} In some cases, UAV moving is implicitly reflected in the analysis. Typical impacts include: 1) UAV moving may affect the CSI acquisition. This is usually reflected by time-correlated channel model \cite{Wang_IWCMC17}, or unknown instantaneous CSI \cite{Xu_Access18}. 2) UAV moving may affect the link reliability, which has been modelled as a Bernoulli process in \cite{Liu_ICUFN17,Liu_JCN_UAV16}. 3) When the UAV is not controlled by the system designer, mobility may cause random variation (constrained by its kinetic parameters, e.g., speed and acceleration rate) in the UAV locations. This is usually described by a random mobility model in the analysis, as that in \cite{Sharma_TWC20,Liang_Access20,Yuan_TIFS19}.

We respectively describe these three cases in more detail. First, UAV mobility causes auto-correlation in time-variant channels. Such a channel model has been considered in \cite{Wang_IWCMC17}, where the auto-correlation coefficient, $\lambda$, is inversely proportional to the UAV speed. Legitimate eavesdropping was investigated therein for an all-UAV network from packet level, and the jamming power, as well as the adaptive modulation and coding (AMC) rate, were jointly optimized during the entire flying period (for all time slots) to maximize the successfully eavesdropped data amount at the legitimate UAV Eve. 

The fast moving of UAV may render the obtained CSI estimation being outdated, i.e., the channel aging effect \cite{Zhang_WC19}. As stated in \cite{Zhang_WC19}, when the UAV is not controlled by the system designer and moves randomly, the auto-correlation property in time would be time-variant, too, which is difficult to measure in practice. As a result, it is in general not easy to predict the channel conditions based on previous observations. In such scenarios, the impact of UAV moving on the propagation links, or the UAV moving itself, is usually described in a stochastic manner.

A commonly adopted model tries to relate the UAV moving to link reliability, which is the second case for implicit trajectory as previously described. In \cite{Liu_ICUFN17,Liu_JCN_UAV16}, the reliability of UAV-related communication links was modelled as a Bernoulli process. When a link breaks down due to UAV moving, the corresponding transmission over this link fails and therefore causes outage. Specifically, with this model, a UAV-aided selective relaying network was investigated in \cite{Liu_ICUFN17}, where a ground control unit  provides backhaul to multiple UAV Tx's, communicating with a ground destination with the aid of multiple UAV relays, in the presence of multiple UAV Eves. Transmission over the backhual links may fail due to the moving of UAV Tx, where the failure probability is modelled with a Bernoulli process. On the other hand, all Eves are colluding to perform MRC of their eavesdropped signals; however, whether an Eve can participate in the MRC depends on its correspond eavesdropping link reliability, which is also modelled as a Bernoulli process. SOP of this system was investigated, and the number of UAV Tx's and relays on the secrecy performance has been discussed via simulations. A similar system model was considered in \cite{Liu_JCN_UAV16}, while the more general Nakagami-$m$ fading was considered in the analysis.

The other way is to describe the position of a randomly moving UAV with physical mobility models. With such models, the position of a UAV is randomly updated in every slot, jointly considering the probability that the UAV stays still at the current waypoint, as well as its speed constraint in certain directions. In \cite{Sharma_TWC20}, UAV relaying was adopted to aid the communication in a hybrid satellite-terrestrial network, in the presence of a UAV Eve. The UAV positions are described with a stochastic mixed mobility model, where the vertical movement of UAV is described by a random waypoint mobility model (RWPM), which defines the probability that UAV stays at a waypoint; during the staying, a random walk model is used to describe the possible spatial excursion. With this model, the corresponding channel characteristics are described, and SOP is analyzed correspondingly. A simpler random mobility model has been considered in \cite{Liang_Access20}, where the propagation distance of UAV-related links randomly varies within a range, constrained on the UAV velocity and the time slot duration. Therein, UAV acts as a friendly jammer to aid the covert communication between two ground nodes.

A random mobility model taking into account more practical constraints, i.e., the smooth turn (ST) model, was adopted in \cite{Yuan_TIFS19,Yuan2019tcom}. Different from the previous models where the actual trajectory is not concerned, the ST mobility model generate random trajectory with practical smooth turns. With this model, secure communication between a ground Tx-Rx pair in the presence of multiple UAV Eves (with random trajectories) was investigated. This has provided a useful framework for secrecy analysis with uncontrollable or unknown UAV Eves. The ST mobility model, which is in accordance with the realistic UAV flying condition, makes the corresponding analysis being of great importance for practical designs. The implicit trajectory-related works are summarized in Table \ref{tab:addlabe3}.

\begin{table*}[tbp]
\tiny
  \caption{UAV-PLS works with implicit or random trajectory.}
  \centering
    \begin{threeparttable}
    \footnotesize
  \resizebox{\textwidth}{!}{
    \begin{tabular}{|c|c|c|c|c|c|c|c|c|c|}
    \hline
    \multicolumn{1}{|c|}{\multirow{2}[4]{*}{Ref.}} & \multicolumn{1}{c|}{\multirow{2}[4]{*}{Scenario}} & \multicolumn{2}{c|}{UAV} & \multicolumn{2}{c|}{Malicious nodes} & \multicolumn{1}{c|}{\multirow{2}[4]{*}{Channel characteristics}} & \multicolumn{1}{c|}{\multirow{2}[4]{*}{Metric}} & \multicolumn{1}{c|}{\multirow{2}[4]{*}{Objective}} & \multicolumn{1}{c|}{\multirow{2}[4]{*}{Mobility reflected by}}\\
\cline{3-6}          &       & \multicolumn{1}{c|}{Role} & \multicolumn{1}{c|}{\tabincell{c}{{Position or}\\{trajectory}}} & \multicolumn{1}{c|}{Role} & \multicolumn{1}{c|}{\tabincell{c}{{Position or}\\{trajectory}}} &       &       &  & \\
    \hline
\cite{Wang_IWCMC17} &\tabincell{c}{{Air: UAV Tx (1), RX (1),}\\{and Eve (1)}\\{({\bf Legitimate eavesdropping})}}& \tabincell{c}{{Tx, Rx}\\{(Suspicious)}\\{Eve}\\{(Legitimate)}}& *\tnote{1} & \tabincell{c}{{Suspicious nodes}\\{(UAV)}}& * &\tabincell{c}{{Normal PL model}\\{Time-correlated channel gain}}&\tabincell{c}{{Data amount}\\{eavesdropped}}& \tabincell{c}{{Optimize jamming power,}\\{and AMC rate}\\{(in all slots)}}& Time correlation in channel\\
\hline
\cite{Liu_ICUFN17,Liu_JCN_UAV16} & \tabincell{c}{{Air: UAV Tx ($M$), Relay ($N$),}\\{UAV Eve ($K$)}\\{Ground: GCU (1), Rx (1)}}& \tabincell{c}{{Tx, DF relay}\\{Eve}}& -- & \tabincell{c}{{Eve}\\{(UAV, colluding)}}& -- & \tabincell{c}{{NLoS}\\{Rayleigh fading \cite{Liu_ICUFN17}}\\{Nakagami-$m$ fading \cite{Liu_JCN_UAV16}}}& SOP & Performance analysis & \tabincell{c}{{Link reliability}\\{(Bernoulli process)}} \\
\hline
\cite{Sharma_TWC20} & \tabincell{c}{{Air: UAV relay ($M$), Eve (1)}\\{Ground: Rx (1)}\\{Space: Satellite (1)}\\{({\bf Satellite network})}}&\tabincell{c}{{DF relay}\\{Eve}}& Random & \tabincell{c}{{Eve}\\{(UAV)}}& Random &\tabincell{c}{{FSPL}\\{Nakagami-$m$ fading}}& -- & -- & Stochastic mobility model\\
\hline
\cite{Liang_Access20} &\tabincell{c}{{Air: UAV (1)}\\{Ground: Alice (1), Bob (1),}\\{Dave (1)}\\{({\bf Covert communications})}}& Friendly jammer & -- & Detector & Fixed & \tabincell{c}{{Normal PL model}\\{Nakagami-$m$/AWGN}}& Privacy rate& -- & \tabincell{c}{{Random variation}\\{in link distances}}\\
\hline
\cite{Yuan_TIFS19} &\tabincell{c}{{Air: UAV Eves ($N$)}\\{Ground: Tx (1), Rx (1)}}& Eve & -- & \tabincell{c}{{Eve}\\{(UAV, colluding)}}& Random & \tabincell{c}{{LoS}\\{Rician fading}}& Secrecy capacity & -- & ST mobility model\\
  \hline   
  \cite{Yuan2019tcom}&\tabincell{c}{{Air: UAV (1)}\\{Ground: Tx (1), Rx (1)} }& \tabincell{c}{Eve}& \tabincell{c}{{ST Mobility} \\{ Model}} & UAV & \tabincell{c}{{ST Mobility} \\{ Model}} & \tabincell{c}{{Rician with {$K > 0$} (A2G)}\\{Rician with {$K \geqslant 0$} (G2G)}}& -- &{--}&--\\
\hline
    \end{tabular}}
\begin{tablenotes}
  \item[1] For brevity, we use ``*'' in this table to describe the case that the trajectory is not explicitly described, or the trajectory does not affect the analysis.
\end{tablenotes}
\end{threeparttable}
  \label{tab:addlabe3}%
\end{table*}

\subsubsection{\textbf{Fixed Trajectory}}

When the UAV trajectory is not controllable by the communication designer, it is considered as a fixed condition, under which the secrecy performance of the system mainly depends on properly-designed resource allocation. As an example, the transmission power can be adaptively increased when the UAV flies closer to the legitimate receiver, and reduced when the UAV is relatively far away from the destination. Such a trajectory-based water-filling-like approach can allocate power more efficiently for the legitimate communication, whereby it is possible to enhance the secrecy rate performance of the system. However, depending on the behaviors of the adversaries, more sophisticated transmission schemes might should be designed.

As typical application scenarios, the UAV may fly over ground facilities, e.g., highway, street, or power line, gathering sensing information and transmit it to a target receiver; or it may be scheduled to fly from a pre-determined take-off point to a landing site. For these scenarios, a fixed linear trajectory is usually assumed \cite{Pan_TWC_UAV209,Shu_PhyCom_2019,Lu_SCIC_UAV34}. In \cite{Pan_TWC_UAV209}, SOP analysis has been conducted jointly considering the geometric positions of the UAV, as well as random location distribution of Eves, while in \cite{Shu_PhyCom_2019}, beamforming and power allocation at a multi-antenna ground Tx has been investigated. In \cite{Lu_SCIC_UAV34}, a FD multi-antenna UAV Rx following linear trajectory is considered, where joint AN projection matrix and power allocation design was investigated to maximize the SR.

In the presence of multiple randomly distributed ground users, circular trajectory is another commonly-adopted assumption. The UAV may patrol following a pre-defined circular trajectory over the area that to be covered, providing service to the users distributed below \cite{chen2020secure33,chen2020power32}. Providing that the trajectory is fixed and known in every transmission slot, the design in general has no obvious difference in the methodology. However, note that new parameters may affect the system performance, e.g., the radius of the circular trajectory might should be designed according to the distribution area of the ground users.

For multiple users with fixed locations on the ground, data transmission can be provided by a UAV Tx in a ``fly--hover--transmit'' manner, in scenarios where the data requirements of users are not time-sensitive. Specifically, the UAV Tx may hover right above one user for its data transmission. Upon accomplishment, the UAV flies to the next user following a fixed trajectory, and this process continues for all users in a sequential order. For such a scenario, UAV swarm-enabled secure communications have been investigated in \cite{Wang-Access_UAV93,wang2019power}. In these works, an Eve is assumed to follow the projection of UAV trajectory on the ground, and large-scale CSI-based power allocation method was proposed to maximize the secrecy throughput in a whole-trajectory-oriented manner.

For the above-mentioned works, in most cases, fixed trajectory implies that the time-varying coordinates of UAV (see $\mathbf{q}_{\rm U}[n]$ previously described in \eqref{eq:12}) are no longer optimization variables, but being considered as fixed parameters in the optimization. However, when the UAV works as a relay, the fixed trajectory may still has impact on the constraints for optimization. For example, consider that the UAV relay flies along a fixed trajectory from the source to destination, and working in a ``store--carry--forward'' manner \cite{Sun2019-UAV56,WangWCL2017-UAV135}. In this case, the trajectory implicitly affects the information-causality of the system. That is, in any time slot, the  accumulated data amount transmitted in the downlink (from relay to destination) should not exceed the accumulated data amount received in the uplink (from source to relay). As the UAV trajectory will directly affect the channel strength in every time slot, different power allocation schemes should be applied to guarantee the information-causality constraint for different trajectories. 

Other mobility model for the UAV relay has been considered. In \cite{abdalla2020securing}, the UAV relay works in an accompany manner, i.e., it follows the trajectory of the ground device. 
Such accompanying mobility model is also suitable for the scenario of wireless surveillance/proactive eavesdropping, where a legitimate UAV Eve usually needs to follow the trajectory of suspicious users for effective surveillance. In \cite{Zhang2019icc}, a legitimate UAV Eve was dispatched to overhear the suspicious communication between two UAV Tx/Rx. Meanwhile with eavesdropping, the legitimate UAV sends jamming signals to degrade the suspicious communications. The jamming power is allocated in the time domain during the flight, which, is directly affected by the accompanying trajectory. A similar scenario was considered in \cite{Zhangm2019ict}, where differently, the legitimate UAV may also work as a spoofing relay depending on its decoding status during the flight trajectory. More practical tracking models have been discussed in \cite{Li_TVT19_Sur}, where the locations of the suspicious nodes cannot be known a priori. The legitimate UAV Eve needs to track the trajectory of the suspicious nodes, via either signal-based or vision-based methods. In this case, the specific tracking methods should be considered in the evaluation of system performance.

\begin{table*}[tbp]
	
\tiny
  \caption{UAV-PLS works with fixed trajectory.}
  \centering
    \begin{threeparttable}
    \footnotesize
  \resizebox{\textwidth}{!}{
    \begin{tabular}{|c|c|c|c|c|c|c|c|c|}
    \hline
    \multicolumn{1}{|c|}{\multirow{2}[4]{*}{Ref.}} & \multicolumn{1}{c|}{\multirow{2}[4]{*}{Scenario}} & \multicolumn{2}{c|}{UAV} & \multicolumn{2}{c|}{Malicious nodes} & \multicolumn{1}{c|}{\multirow{2}[4]{*}{Channel characteristics}} & \multicolumn{1}{c|}{\multirow{2}[4]{*}{Metric}} & \multicolumn{1}{c|}{\multirow{2}[4]{*}{Objective}} \\
\cline{3-6}          &       & \multicolumn{1}{c|}{Role} & \multicolumn{1}{c|}{\tabincell{c}{{Position or}\\{trajectory}}} & \multicolumn{1}{c|}{Role} & \multicolumn{1}{c|}{\tabincell{c}{{Position or}\\{trajectory}}} &       &       &  \\

     \hline
   \cite{Lu_SCIC_UAV34}& \tabincell{c}{{Air: UAV (1)}\\{Ground: Alice (1), Eve (1)}\\{(\textbf{Multi-antenna FD UAV})}}& \tabincell{c}{Bob} & Fixed/Straight line & Eve & \tabincell{c}{Fixed, known}& \tabincell{c}{{LoS, no fading}}& SR & \tabincell{c}{{Optimize beamforming, AN matrix,}\\{ and power allocation}}\\
   
   \hline 
    \cite{Pan_TWC_UAV209}& \tabincell{c}{{Air: Legitimate UAV (1),}\\{ UAV Eve (1)}\\{Ground: Ground station (1)}\\({\textbf{Power-line inspection}})}& \tabincell{c}{{Tx/Rx,}\\{Eve}}& -- & \tabincell{c}{{Eve}\\(UAV)}& Randomly distributed &\tabincell{c}{{LoS, Rayleigh fading}}& SOP& Performance analysis\\
    
        \hline  
    \cite{Shu_PhyCom_2019}&\tabincell{c}{{Air: UAV (1)}\\{Ground: Alice (1), Eve (1)}\\{({\bf Directional modulation})}}& Bob & -- & Eve& Fixed, known & \tabincell{c}{{LoS, no fading}}& Secrecy sum-rate&\tabincell{c}{{Optimize beamformer,}\\{and AN power allocation}}\\
    
     \hline     
     \cite{chen2020secure33,chen2020power32}&\tabincell{c}{{Air: UAV BS (1)}\\{Ground: Common users ($K$)} \\ {Secure user (1), Eve(1)} \\ {({\bf NOMA})}}& \tabincell{c}{{Tx}}& \tabincell{c}{Fixed/Circular} & -- & -- & \tabincell{c}{Rician fading}& \tabincell{c}{{Sum rate}\\{(SR constraint)}}&\tabincell{c}{Power allocation}\\
    
    \hline
    \cite{Wang-Access_UAV93}& \tabincell{c}{{Air: UAV (M),}\\{Ground: Alice (M), Eve (1)}}& --& Fixed & -- & Unknown& \tabincell{c}{{Probabilistic LoS/NLoS,}\\{Rayleigh fading}}& Secrecy throughput& \tabincell{c}{{Optimize confidential message power.}\\{AN power, and transmission duration}}\\

    \hline    
    \cite{wang2019power}&\tabincell{c}{{Air: UAVs (L)}\\{Ground: LUs (K)} \\ {Eve(1)} \\{(\textbf{multi-antenna Users and Eve})}}& \tabincell{c}{{--}}& \tabincell{c}{--} & -- & \tabincell{c}{{randomly moving} \\ {along UAVs}} & \tabincell{c}{{LOS/NLOS}}& {--}&\tabincell{c}{{Optimize UAVs' transmit power} \\{and transmission duration scheduling}}\\

  \hline
  \cite{WangWCL2017-UAV135} & \tabincell{c}{{Air: UAV (1)}\\{Ground: Alice (1), Bob (1),}\\{ Eve (1)}}& Relay & --&  Eve & \tabincell{c}{{Fixed, known}}& FSPL & ASR & Optimize UAV's transmit power \\ 
  \hline
  \cite{Sun2019-UAV56}&\tabincell{c}{{Air: UAV (1)}\\{Ground: Alice (1), Bob (1),}\\{ Eve (M)}}& --& -- &-- & \tabincell{c}{{Partially known} \\{in a certain area}}&--& SR& \tabincell{c}{{Optimize source's/relay's}\\{transmit power}}\\
   \hline     
   \cite{abdalla2020securing}&\tabincell{c}{{Air: UAV relay (1)}\\{Ground: BS (1), IoT device (1), Eve(1)} }& \tabincell{c}{{--}}& \tabincell{c}{Follow IoT device} & -- & \tabincell{c}{{Fixed, known}} & \tabincell{c}{{Probabilistic LOS/NLOS} \\ {Nakagami-$m$}}& {--}&\tabincell{c}{Performance analysis}\\


   \hline    
   \cite{lu2020proactive}&\tabincell{c}{{Air: UAV relay(1)}\\{Ground: Tx (1), Rx (1)} \\{monitor (1)}\\ {(\textbf{Surveillance})}}& \tabincell{c}{{AF/DF relay} \\ {jammer} \\ {(FD)}}& Fixed & Suspicious Tx & Fixed & \tabincell{c}{{LoS, no fading (A2G)}\\{NLoS, Rayleigh (G2G)}}& \tabincell{c}{{Average} \\{eavesdropping rate}} &\tabincell{c}{{Jamming power allocation }}\\
   
   \hline          
     \cite{Li_TVT19_Sur}&\tabincell{c}{{Air: Legitimate UAV (1)}\\{UAV Tx (1), UAV Rx (1)} \\ {(\textbf{Surveillance})}}& \tabincell{c}{{Tx/Rx/jammer}}& \tabincell{c}{{Tracking Suspicious} \\ {UAV's trajectory}} & -- & * & \tabincell{c}{{LoS, no fading}}& Eavesdropped packets &\tabincell{c}{{--}}\\
     \hline  
     \cite{Zhang2019icc}&\tabincell{c}{{Air: Legitimate UAV (1)}\\{UAV Tx (1), UAV Rx (1)} \\ {(\textbf{Surveillance})}}& --& \tabincell{c}{Predetermined} & -- & {Fixed} & \tabincell{c}{{--}}& Eavesdropping Rate &\tabincell{c}{{--}}\\
     \hline  
     \cite{Zhangm2019ict}&\tabincell{c}{{Air: Legitimate UAV (1)}\\{UAV Tx (1), UAV Rx (1)} \\ {(\textbf{Surveillance})}}& \tabincell{c}{{Tx/Rx} \\{AF relay/jammer}\\{(FD)}}& \tabincell{c}{--} & -- & {--} & \tabincell{c}{{--}}& -- &\tabincell{c}{{--}}\\

     \hline
    \end{tabular}}
\end{threeparttable}
  \label{tab:fixed_traj}%
\end{table*}

\begin{table*}[tbp]
\tiny
  \caption{Trajectory optimization part I: UAV as Tx/Rx.}
  \centering
    \begin{threeparttable}
    \footnotesize
  \resizebox{\textwidth}{!}{
    \begin{tabular}{|c|c|c|c|c|c|c|c|c|}
    \hline
    \multicolumn{1}{|c|}{\multirow{2}[4]{*}{Ref.}} & \multicolumn{1}{c|}{\multirow{2}[4]{*}{Scenario}} & \multicolumn{2}{c|}{UAV} & \multicolumn{2}{c|}{Malicious nodes} & \multicolumn{1}{c|}{\multirow{2}[4]{*}{Channel characteristics}} & \multicolumn{1}{c|}{\multirow{2}[4]{*}{Metric}} & \multicolumn{1}{c|}{\multirow{2}[4]{*}{Objective}} \\
\cline{3-6}          &       & \multicolumn{1}{c|}{Role} & \multicolumn{1}{c|}{\tabincell{c}{{Position or}\\{trajectory}}} & \multicolumn{1}{c|}{Role} &
 \multicolumn{1}{c|}{\tabincell{c}{{Position or}\\{trajectory}}} &       &       &  \\
 
   
    \hline

    \cite{ZhangWCNC2017} &\tabincell{c}{{Air: UAV  Tx (1)}\\{Ground: Bob (1),} \\ {and Eve(1)}}& \tabincell{c}{Tx} & \tabincell{c}{{To be optimized} \\ {(Fixed altitude)}} & Eve & \tabincell{c}{{Fixed} \\ {Known}} & LoS, no fading & ASR & \tabincell{c}{{Optimize UAV's trajectory}\\{ and transmit power}}\\     
    \hline 
    
    \cite{Gao2019-UAV-80} &-- & \tabincell{c}{{--}} & \tabincell{c}{{--} \\ {(no fly zone constraint)}} & \tabincell{c}{{--} } & -- & -- & -- & --\\     
    \hline 
    
    \cite{Wang2018-UAV18} &-- & -- & \tabincell{c}{{To be optimized} \\ {(Fixed altitude)} } & -- & -- & -- & \tabincell{c}{{SEE}} & \tabincell{c}{{Optimize UAV's trajectory,} \\ {velocity, and acceleration}}  \\     
    \hline  
    
    \cite{Zhang2019-UAV50} &\tabincell{c}{{Air: UAV  (1)}\\{Ground: Legitimate node (1),} \\ {and Eve(1)}\\{({\bf Uplink and downlink})}}& Tx/Rx & -- & -- & -- & \tabincell{c}{ {LoS, no fading (G2A, A2G)} \\ {NloS, Rayleigh (G2G)} } & ASR & \tabincell{c}{{Optimize UAV's trajectory}\\{ and transmit power}}\\     
    \hline 
    
	\cite{Fang2020wclirs} &\tabincell{c}{{Air: UAV Tx (1)}  \\ {Ground: Bob (1), IRS (1)} \\ {and Eve(1)} \\ {(\textbf{IRS})}}& \tabincell{c}{ {Tx} } & -- & -- & -- & \tabincell{c}{ {LoS, no fading (A2G)} \\ {NLoS, Rayleigh (G2G)} } & -- & \tabincell{c}{{Optimize UAV's trajectory,} \\ {transmit power,} \\{and IRS phase shifts} }  \\ 
    \hline 
    
    \cite{Wang2020tvt} &\tabincell{c}{{Air: UAV Tx (1),}  \\ {Ground: Bob (1) and Eve (1)} \\ {(\textbf{Finite blocklength})}}& -- & -- & -- & -- & \tabincell{c}{ LoS, no fading } & \tabincell{c}{{ASR modified by}\\{decoding error probability}\\{in finite blocklength}} & \tabincell{c}{{Optimize UAV's trajectory,} \\ { and transmit power}}  \\ 
    \hline 
    
    \cite{Li-ICL-UAV32} &\tabincell{c}{{Air: UAV Tx (1) } \\ {Ground: Legitimate nodes ($K$)} \\ {and Eve (1)}}& -- & -- & -- & -- & -- & \tabincell{c}{{Worst user ASR}} & \tabincell{c}{{Optimize UAV's trajectory, } \\ { transmit power, }\\{ and user scheduling}}\\     
    \hline

    \cite{Wu_ICASSP_UAV165} &\tabincell{c}{{--}\\{({\bf Zipf content distribution})}}& -- & -- & -- & -- & -- & \tabincell{c}{{Transmission delay}\\{(Secure QoS constraint)}} & \tabincell{c}{{Optimize UAV's trajectory,} \\ {and user association }}  \\     
    \hline   
    
     \cite{Cai2019-UAV97} &\tabincell{c}{{--} \\ {(\textbf{OFDMA})}}& -- & -- & -- & \tabincell{c}{{--} \\ {(with error)} } & -- & \tabincell{c}{{EE}\\{(Eve's SNR constraint)}\\{(Flight power considered)}} & \tabincell{c}{{Optimize UAV's trajectory,}\\{ transmit power, velocity,} \\ {and user scheduling}}\\     
    \hline

    \cite{Cui2018-UAV144} &\tabincell{c}{{Air: UAV  Tx (1)}\\{Ground: Bob (1),} \\ {and Eves ($K$)}}& -- & -- &\tabincell{c}{{Eves}\\{(non-colluding)}} & -- &-- & \tabincell{c}{{worst-case ASR} } & \tabincell{c}{{Optimize UAV's trajectory}\\{ and transmit power}}\\     
    \hline

    \cite{sun2021iot} &\tabincell{c}{{--} \\ {(\textbf{mmWave, SWIPT})}  \\ {(\textbf{Multi-antenna UAV})}}& -- & -- & \tabincell{c}{{colluding Eve} } & -- & -- & ASR & \tabincell{c}{{Optimize UAV's trajectory,} \\ {transmit power, } \\ {and power splitting ratio}}\\     
    \hline

    \cite{Savkin2020wcl} &\tabincell{c}{{--} \\ {(\textbf{Stationary/Mobile Bob and Eves})}\\{(\textbf{3D UAV motion model})} }& -- & \tabincell{c}{{To be optimized}\\{(3D)} } & -- & \tabincell{c}{{Fixed or mobile} \\ {(can be known)} }& \tabincell{c}{ {Path loss + Rician fading} } & \tabincell{c}{{Control energy} \\ {(Bob/Eve's Rate constraints)}} & \tabincell{c}{{Optimize UAV's horizontal,} \\ {angular, and vertical speeds} }\\     
    \hline 
    
    \cite{Hong2019access} &\tabincell{c}{{Air: UAV Tx (1),}  \\ {Ground: Information Rx (1),}  \\ { and Energy Rx ($K$)} \\ {(\textbf{SWIFT})}}& SWIFT Tx & \tabincell{c}{{To be optimized} \\ {(Fixed altitude)} } & \tabincell{c}{{Energy Rx}} & \tabincell{c}{{Fixed} \\ {known} } & \tabincell{c}{ {LoS, no fading} } & \tabincell{c}{{worst-case ASR}\\{(ER energy constraint)}} & \tabincell{c}{{Optimize UAV's trajectory,} \\{ and transmit power }}  \\     
%
    \hline 

    \cite{wang2020tcom} &\tabincell{c}{{Air: UAV BS (1),} \\ {Ground: Secrecy-required Users (SU) ($K$)}\\ {and QoS-required Users (QU) ($M$)} \\ {(\textbf{NOMA})}}& Tx & -- & \tabincell{c}{{Unpaired QUs} } & --& -- & \tabincell{c}{{Worst user ASR}} & \tabincell{c}{{Optimize UAV's trajectory,} \\ {transmit power, } \\ {and user pairing} }\\     
    \hline    
    
    \cite{Yao2019-UAV61,Yao2020twc} &\tabincell{c}{{Air: UAV Tx (1),} \\ {Ground: Rx ($K$),}  \\ {and Eves ($J$)} \\ {(\textbf{CoMP reception})}}& -- & \tabincell{c}{{To be optimized}\\{(3D)}} & \tabincell{c}{{Colluding/non-colluding}\\{ Eves}} & -- & -- & ASR & \tabincell{c}{{Optimize UAV's trajectory,} \\ {and transmit power}}\\     
%
    \hline  

	\cite{Li2020wcl} &\tabincell{c}{{Air: UAV Tx (2),}  \\ {Ground: Destinations (2),}  \\ { and Eve (1)} \\ {(\textbf{Interference channel})}}& \tabincell{c}{ {Tx} \\ {(spectrum sharing)}} & -- & \tabincell{c}{{Eve} \\ {wiretap both links}} & -- & \tabincell{c}{ { LoS, Rician} } & \tabincell{c}{{ASR}\\{(Sum of two links)}} & \tabincell{c}{{--}\\{(of two UAVs)}}  \\  
	
%
    \hline

     \cite{Zhou_TC_UAV158} &\tabincell{c}{{Air: UAV Tx (1),} \\ {Ground: Secondary Rx (1),}  \\ {Primary Rx ($L$), Eves ($K$)} \\ {(\textbf{Cognitive Radio})}}& \tabincell{c}{ Secondary Tx } & \tabincell{c}{{To be optimized} \\ {(Fixed altitude)} } & \tabincell{c}{{Eves}\\{(non-colluding)}} & \tabincell{c}{{Fixed} \\ {(known with error)}} & \tabincell{c}{ {LoS, no fading} } & \tabincell{c}{{worst-case ASR}} & --  \\    
     
         \hline     
     
    \cite{Tang2020access} &\tabincell{c}{{Air: UAV Tx (1),}  \\ {Ground: Secondary Rx ($K$)} \\ {primary Rx ($I$), Eve(1)} \\ {(\textbf{Cognitive radio, NOMA})}}& -- & \tabincell{c}{{To be optimized}\\{(3D)} } & -- & \tabincell{c}{{Fixed} \\ {known} } & -- & -- & --  \\ 
    
        \hline

    \cite{wang2020establishing}&\tabincell{c}{{Air: UAV Tx (1)}\\{Ground: Users ($K$), Eves ($N$)}\\{(\textbf{Virtual BS via UAV})}}& \tabincell{c}{{Tx}}& \tabincell{c}{{To be optimized}\\{(Fixed altitude)}} & -- & -- & -- & \tabincell{c}{{Radius of the virtual BS} \\ {(disk covering problem)}}&\tabincell{c}{{Virtual BS placement}\\ {and visiting order}}\\
%

   \hline    
     \cite{liu2020multi17}&\tabincell{c}{{Air: UAVs ($M$)}\\{Ground: Users ($K$), unauthorized user (1)} }& -- & -- & \tabincell{c}{{Eve}\\{(unauthorized user)}} & -- & -- & \tabincell{c}{{Worst-user ASR}}&\tabincell{c}{{Optimize UAV's}\\ {trajectory, velocity, } \\{ acceleration, transit power}\\{and users scheduling}}\\

    \hline  
    
    \cite{Zhou2019globec,zhou2019uav} &\tabincell{c}{{Air: UAV (1)}\\{Ground: Users ($K$)} \\ {(\textbf{Convert communications})}}& \tabincell{c}{ {Rx/Tx}\\{(FD UAV)} } & \tabincell{c}{{--} } & \tabincell{c}{{Willie}\\{(Unscheduled users)}} & \tabincell{c}{{--} } & \tabincell{c}{ {LoS, no fading (A2G/G2A)}\\{Rayleigh (Self-interference channel)} } & \tabincell{c}{{Worst-user rate}\\{(Covertness constraint)}} & \tabincell{c}{{Optimize UAV's trajectory,} \\{ AN power, user scheduling} }  \\     
    \hline

    \cite{Zhou_TSP19} &\tabincell{c}{{Air: UAV Tx (1),} \\ {Ground: Bob (1), Willie (1)} \\ {(\textbf{Convert communications})}}& \tabincell{c}{ {Tx}} & -- & \tabincell{c}{{Willie}} & \tabincell{c}{{Fixed} \\ {(known with error)}} & \tabincell{c}{ {LoS, no fading} } & \tabincell{c}{{Average rate} \\ {(Covertness constraint)} } & \tabincell{c}{{Optimize UAV's trajectory,} \\ {and transmit power}}  \\  
       
    \hline

\end{tabular}}
\end{threeparttable}
  \label{tab:traj_opt}%
\end{table*}

\begin{table*}[tbp]
\tiny
  \caption{Trajectory optimization part II: UAV as relay, jammer, and dual-UAV case.}
  \centering
    \begin{threeparttable}
    \footnotesize
  \resizebox{\textwidth}{!}{
    \begin{tabular}{|c|c|c|c|c|c|c|c|c|}
    \hline
    \multicolumn{1}{|c|}{\multirow{2}[4]{*}{Ref.}} & \multicolumn{1}{c|}{\multirow{2}[4]{*}{Scenario}} & \multicolumn{2}{c|}{UAV} & \multicolumn{2}{c|}{Malicious nodes} & \multicolumn{1}{c|}{\multirow{2}[4]{*}{Channel characteristics}} & \multicolumn{1}{c|}{\multirow{2}[4]{*}{Metric}} & \multicolumn{1}{c|}{\multirow{2}[4]{*}{Objective}} \\
\cline{3-6}          &       & \multicolumn{1}{c|}{Role} & \multicolumn{1}{c|}{\tabincell{c}{{Position or}\\{trajectory}}} & \multicolumn{1}{c|}{Role} &
 \multicolumn{1}{c|}{\tabincell{c}{{Position or}\\{trajectory}}} &       &       &  \\
    
    \hline

    \cite{SunGlobecom2018-UAV42} &\tabincell{c}{{Air: UAV relay (1)}\\ {Ground: Alice (1),} \\ {Bob (1), and Eve (1)}}& \tabincell{c}{{DF relay}} & \tabincell{c}{{To be optimized}\\{(3D)}} & Eve & \tabincell{c}{{Fixed}\\{known}} & LoS, no fading & worst-case ASR & \tabincell{c}{{Optimize UAV's trajectory,} \\ {bandwith,} \\ {and Alice/UAV transmit power}}\\     
    \hline
    
    \cite{Xiao-TGCN-UAV44} &--& -- & \tabincell{c}{{To be optimized} \\ {(Fixed altitude)}} & -- & -- & \tabincell{c}{ {LoS, no fading (G2A, A2G)} \\ {NloS, Rayleigh (G2G)} } & \tabincell{c}{{SEE}} & \tabincell{c}{{Optimize UAV's trajectory, } \\ { velocity, acceleration, and}\\{ UAV/Source transmit power}}\\     
    \hline     
    
     \cite{Wang2018-UAV20} &--& \tabincell{c}{{--} \\ {(with buffer) }\\{(Information-causality)}} & -- & -- & -- & \tabincell{c}{ {LoS, no fading}} & \tabincell{c}{ASR} & \tabincell{c}{{Optimize UAV's trajectory}\\{ and Source/UAV's transmit power}}\\     
    \hline

    \cite{ShenVTC-UAV55,Shensenseors-UAV55} &--& -- & -- & -- & \tabincell{c}{{Distributed} \\ {on a line} } & --& \tabincell{c}{{ESR}\\{(Averaged over}\\{ possible Eve's locations)}} & \tabincell{c}{{Optimize UAV's trajectory} }\\     

    \hline

    \cite{ SunICC2019-UAV52 } &--& -- & -- & \tabincell{c}{{Eve}\\{(user with}\\{lower security level)} } & \tabincell{c}{{--} \\ {(with error)}} & -- & \tabincell{c}{{ASR} \\ {(Eve's rate constraint)}} & \tabincell{c}{{Optimize UAV's trajectory,} \\ {Alice/UAV transmit power,} \\ {and time scheduling}}  \\ 

    \hline   
    
    \cite{TC-Cheng-UAV4, Cheng-ICC-UAV38} &\tabincell{c}{{Air: UAV relay (1) } \\  {Ground: Legitimate users (M), } \\ {BS (1), and Eve (1)}\\{(\textbf{Caching users/UAV})}}& \tabincell{c}{{--}\\{(with caching)}} & -- & -- & \tabincell{c}{{Fixed} \\ {Known}  } & -- & \tabincell{c}{{Minimum ASR}} & \tabincell{c}{{Optimize UAVs' trajectory,} \\ {and time scheduling}}\\

   \hline   
    
     \cite{abdalla2020securing1}&\tabincell{c}{{Air: UAV Relay (1)}\\{Ground: BS (1), Users ($M$), Eves ($K$)}\\{(\textbf{Mobile users})}}& --& -- & \tabincell{c}{{Eves}\\{(non-colluding)}} & -- & -- & \tabincell{c}{{Worst-case ASR}}&\tabincell{c}{{Optimize UAV's trajectory}\\ {and transmit power}}\\ 
 
     \hline
        
    \cite{Ahmed2019-UAV140} &\tabincell{c}{{Air: UAV relay (1)}\\ {Ground: Legitimate users ($U$),} \\ {BS (1), and Eves ($E$)} \\ {(\textbf{Information dissemination})}}& \tabincell{c}{{--}\\{(Information-causality)}} & -- & -- & -- & -- & \tabincell{c}{{worst-case ASR}} & \tabincell{c}{{Optimize UAV's trajectory,} \\ {and BS/UAV transmit power}}\\     
    
    \hline    
    
    \cite{ahmed2020joint} &\tabincell{c}{{Air: UAV Relay(1),} \\ {Ground:BS (1), User (1), Eves ($M$)}  }& -- & -- & -- & \tabincell{c}{{Distributed in a} \\ {uncertainty region}} & -- & \tabincell{c}{{Worst-case SEE} } & \tabincell{c}{{Optimize UAV's trajectory} \\ {speed, acceleration,} \\ {and UAV's/ BS's power} }  \\ 
    
    \hline

    \cite{Li_ICL_UA138} &\tabincell{c}{{Air: UAV jammer (1)}\\{Ground: Source (1),} \\ {Destination (1), and Eve (1)}}& Friendly jammer & -- & -- & \tabincell{c}{{Fixed} \\ {Known}} & \tabincell{c}{ {LoS, no fading (A2G)} \\ {NloS, Rayleigh (G2G)} }& ASR & \tabincell{c}{{Optimize UAV's trajectory,}\\{ and UAV/Source transmit power}}\\     
    \hline 

    \cite{Nnamani2020access} &--& -- & -- & -- & \tabincell{c}{{Distributed in a} \\ {uncertainty region}} & -- & \tabincell{c}{{ESR}\\{(averaged over Eve's}\\{channel distribution)}} & --  \\     
    \hline          
    
    \cite{  Roh2019milcom} &\tabincell{c}{{--}\\{(\textbf{Location error model}}\\{\textbf{for Bob/Eve})}}& -- & -- & -- & \tabincell{c}{{Fixed}  \\ {(known with error)}} & -- & \tabincell{c}{{worst-case ASR}\\{(worst over location errors)}} & \tabincell{c}{{Optimize UAV's trajectory,} \\ {and transmit power }}  \\ 

   \hline  
      
    \cite{nguyen2020uav}&\tabincell{c}{{Air: UAV jammer (1)}\\{Ground: Secondary Tx/Rx (1/1)} \\ {Primary Rx (1), Eve (1)} \\{(\textbf{Cognitive radio})} }& -- & \tabincell{c}{{To be optimized}\\{(3D)}} & -- & \tabincell{c}{{Fixed} \\ {(Perfect/imperfectly known)}} & -- & \tabincell{c}{{ASR/Worst-case ASR}}&\tabincell{c}{{Optimize UAV's trajectory}\\ { and UAV's/ST's power}}\\    
        
    \hline         
    
    \cite{Xu2021tcom} &\tabincell{c}{{Air: UAV (1),} \\ {Ground: Bob (1), Eve (1)} }& \tabincell{c}{{Tx/jammer} \\ {({via power splitting})} } & \tabincell{c}{{To be optimized}\\{(Fixed altitude)}} & -- & \tabincell{c}{{Fixed} \\ {Known}}  & \tabincell{c}{ {LoS, no fading} } & \tabincell{c}{{ASR}} & \tabincell{c}{{Optimize UAV's trajectory,} \\ {transmit power and,} \\ {power splitting ratio}}\\   
    
       \hline   
   
     \cite{li2020uav}&\tabincell{c}{{Air: UAV (1)}\\{Ground: Users ($K$)}}& --& -- & {Unscheduled users} & -- & --& \tabincell{c}{{worst-case ASR}}&\tabincell{c}{{Optimize UAV's trajectory,}\\ {transmit power splitting,} \\{and user scheduling}}\\
       
    \hline      
    
    \cite{Duo2020tvt} &\tabincell{c}{{Air: UAV (1),}  \\ {Ground: Source (1), Eve(1)} }& \tabincell{c}{ {Rx/jammer} \\ {(FD UAV)}} & -- & Eve & -- & \tabincell{c}{ {LoS, no fading (A2G/G2A)} \\ {NLoS, Rayleigh (G2G)}  } & \tabincell{c}{{SEE}} & \tabincell{c}{{Optimize UAV's trajectory,} \\ {and UAV/source's power}}  \\ 
    \hline  
    
  
    \cite{zhou2020uav} &\tabincell{c}{{Air: UAV (1),}  \\ {Ground: Sensor nodes ($K$)}\\{(\textbf{Data collection})} }& -- & -- & \tabincell{c}{{Eve}\\{(unscheduled users)}} & -- & -- & \tabincell{c}{{worst-case ASR}} & \tabincell{c}{{Optimize UAV's trajectory,} \\{ AN power, rate determination,} \\ {and user scheduling}}  \\     
 
         \hline     

    \cite{Lee_TVT_UAV28} &\tabincell{c}{{Air: UAV Tx (1), UAV jammer (1)} \\ {Ground: Users ($K$)}}& \tabincell{c}{{Tx/Jammer}} & -- & \tabincell{c}{{Eve} \\ {(unscheduled users)}} & -- & \tabincell{c}{ {LoS, no fading}} & \tabincell{c}{ASR} & \tabincell{c}{{Optimize UAVs' trajectory,}\\{ transmit power,} \\ {and user scheduling}}\\     
    \hline

    \cite{Zhong_ICL_UAV24} &\tabincell{c}{{Air: UAV Tx (1), UAV jammer (1)}\\ {Ground: Rx (1), Eve (1)} }& -- & -- & -- & \tabincell{c}{{--} \\ {(with error)} } & -- & -- & \tabincell{c}{{Optimize UAVs' trajectories }\\{ and transmit/jamming power}}\\     
    \hline          
    
    \cite{Mamaghani2020tvt} &\tabincell{c}{{--}\\{(\textbf{Energy harvesting Rx})}}& -- & -- & -- & \tabincell{c}{{Unknown} \\ {(circular estimated}\\{ region)}} & \tabincell{c}{ {Probabilistic LoS/NLoS} } & -- & \tabincell{c}{{Optimize UAVs' trajectory,} \\ {transmit power and,} \\ {Rx's power splitting ratio}}\\     
    \hline              
    
    \cite{li2018mobile} &--& -- & -- & -- & \tabincell{c}{{Fixed} \\ {Known}  } & \tabincell{c}{ {LoS, no fading} } & -- & \tabincell{c}{{Optimize UAVs' trajectories,} \\ {and transmit/jamming power}}\\     
%
    \hline

    \cite{Li_Access_UAV37} &\tabincell{c}{{Air: UAV Tx (1), UAV jammer (1)} \\ {Ground: Rx (1), Eves ($K$)}}& -- & -- & \tabincell{c}{{Eves}\\{(non-colluding)}} & \tabincell{c}{{Fixed} \\ {(known with error)}} & -- & \tabincell{c}{{worst-case ASR}} & --  \\     
    \hline        
    
    \cite{Cai_JASC_UAV9, Cai_ICC_UAV59} &\tabincell{c}{{Air: UAV Tx (1), UAV jammer (1)} \\ {Ground: Rx ($K$), Eves ($N$)} }& -- & -- & -- & -- & -- & -- & \tabincell{c}{{Optimize UAVs' trajectories }\\{ and user scheduling}}\\     
    \hline      
      
    \cite{Zhou_TVT_UAV10} &--& -- & \tabincell{c}{{To be optimized} \\ {(Fixed altitude)} \\ {(collision avoidance)}} & -- & \tabincell{c}{{Fixed} \\ {Known}  } & -- & -- & \tabincell{c}{{Optimize UAVs' trajectories, } \\ { and transmit power}}\\     
%

    \hline       
             
    \cite{Cai2020tcom} &\tabincell{c}{{--} \\ {(\textbf{Multi-antenna UAV jammer, }} \\ {\textbf{OFDMA})}}& -- & \tabincell{c}{{UAV Tx trajectory:} \\ {to be optimized;} \\ {Jammer: Trajectory fixed} } & -- & \tabincell{c}{{Fixed} \\ {(known with error)}} & -- & \tabincell{c}{{EE} \\ {(Eve's SNR constraint)}} & \tabincell{c}{{Optimize Tx UAV's trajectory,} \\{velocity, transmit power,} \\ {and jammer's AN}}  \\ 
 
   \hline     
    \cite{zhang2020uav32,zhang2020multi32}&\tabincell{c}{{Air: UAV BS(1), UAV jammer($J$)}\\{Ground: Users ($N$), Eve(K)} \\ {(\textbf{Multi-antenna UAV})}\\{(\textbf{Reinforcement learning})} }& -- & \tabincell{c}{{To be optimized}\\{(3D)}} & -- & \tabincell{c}{{Fixed} \\ {Known}} & \tabincell{c}{{probabilistic LoS/NLoS}} & \tabincell{c}{{Worst-case ASR}}&\tabincell{c}{{Optimize UAVs' trajectories}\\ { and transmit power }}\\ 
    
        \hline    
    
    \cite{Miao_Access_UAV156} &\tabincell{c}{{Air: UAV relay (1), UAV jammer (1)} \\ {Ground: Source (1), Destination (1), Eve (1)}}& \tabincell{c}{ {Relay}\\{(information-causality)} \\ {friendly jammer}} & \tabincell{c}{{To be optimized} \\ {(Fixed altitude)} } & -- & \tabincell{c}{{Fixed} \\ {known} } & -- & \tabincell{c}{{ASR}} & \tabincell{c}{{Optimize UAVs' trajectories,} \\{ and UAV/Source's transmit power }}  \\  
      
    \hline      
       
    \cite{Hua_TVT_UAV7} &\tabincell{c}{{Air: UAV jammers ($M_1$), UAV Tx's ($M_2$)} \\ {Ground: Users ($K_1$), Eves ($K_2$)}}& \tabincell{c}{{Tx/jammer}} & -- & \tabincell{c}{{Eves}\\{(non-colluding)}} & -- & -- & \tabincell{c}{{worst-case SEE}} & \tabincell{c}{{Optimize UAVs' trajectories, } \\ { transmit power, }\\{ and user scheduling}}\\

    \hline         

    \cite{Zhang2019infocom} &\tabincell{c}{{Air: UAV source (1), UAV Rx (1)} \\ {UAV jammer (1), UAV relays ($N$)} \\ {Ground: Interferer (1), Eve (1)} }& \tabincell{c}{{Tx/Rx/jammer/DF relay }} & \tabincell{c}{{Jammer's trajectory:} \\ {to be optimized;} \\ {others fixed} \\ {(Fixed altitude)}} & -- & \tabincell{c}{{Distributed} \\ {in a region} }& \tabincell{c}{ {LoS, no fading (A2A/G2A)} \\ NLoS, no fading (A2G) } & \tabincell{c}{{Network SR}} & \tabincell{c}{{Optimize UAV jammer's} \\ {trajectory,} \\ {and bandwidth allocation} }\\     
    
    \hline

\end{tabular}}
\end{threeparttable}
  \label{tab:traj_opt}%
\end{table*}

\subsection{Trajectory Optimization}

We now introduce the most important research direction in UAV-related communications, that is, trajectory optimization. There exist quite a large amount of works in this category. To allow the reader find his/her interested scenario conveniently, we structure this subsection according to the UAV's roles, as that has been described previously in Section II.C.

\textit{\textbf{1) UAV Transmitter/Receiver:}}
We start from the basic three-node wiretap model. A Tx attempts to send confidential messages to a legitimate Rx in the presence of an Eve, where the UAV may act as either Tx or Rx. PLS design is slightly different for these two cases: When UAV acts as Tx, its trajectory will affect both the legitimate and the eavesdropping links. Generally, the trajectory should be designed to be close to the legitimate Rx, while trying to stay away from potential Eves (if their positions are known). On the other hand, when UAV acts as a Rx, its trajectory only affects the legitimate link. Nevertheless, by letting the trajectory be closer to the legitimate Tx, lower transmit power is expected, and hence, information leakage in the eavesdropping link can be reduced.

Such a three-node system model has been considered in \cite{ZhangWCNC2017,Gao2019-UAV-80,Wang2018-UAV18,Zhang2019-UAV50}. In \cite{ZhangWCNC2017}, an algorithm was proposed to maximize the SR by jointly optimizing the UAV’s trajectory and transmit power over a finite horizon area. The same design metric and optimization objective were investigated in \cite{Gao2019-UAV-80}, where the authors further added no-fly zone constraints in the analysis. To strike the trade-off between secrecy throughput and the UAV's energy consumption, \cite{Wang2018-UAV18} maximized the SEE while satisfying the general communication performance requirements. These studies mainly focused on the UAV-to-ground (U2G) case. In \cite{Zhang2019-UAV50}, the authors considered both the U2G and ground-to-UAV (G2U) communications. Specifically, ASE was maximized for these two scenarios, where the UAV's trajectory and the transmit power were jointly optimized. 
The three-node model has been considered for some specific scenarios, such as RIS, finite-blocklength communications, mmWave, etc. \cite{Fang2020wclirs,Wang2020tvt,WuICCC-UAV145}. RIS assisted UAV network was considered in \cite{Fang2020wclirs}, where the authors proposed an iterative algorithm to maximize the SR by jointly optimizing the trajectory, the transmission power and the phase shifts of RIS. For UAV-enabled secure communication with finite blocklength, an iteration algorithm was proposed to maximize the ASR in \cite{Wang2020tvt}. By introducing the concept of mobile secrecy guard cone and with the aid of a 3D sectorized antenna model, the authors in \cite{WuICCC-UAV145} investigated secure transmission for the UAV-enabled millimeter wave communication. The ASR was maximized by jointly optimizing the UAV's trajectory and power allocation.

More general multi-node scenarios were investigated in ~\cite{Li-ICL-UAV32, Wu_ICASSP_UAV165, Savkin2020wcl, Cai2019-UAV97, Cui2018-UAV144, Fan2019-UAV143, wang2020establishing, liu2020multi17}. Compared with the three-node model, new optimization targets and variables emerge. In the multi-user scenarios, the minimum SR among users is often taken as the design objective for fairness. Besides, the system usually serves multiple users via TDMA or FDMA protocols, where user scheduling or association can be optimized to further enhance the secrecy performance. Relaxation or penalty methods are often adopted to handle binary variables about user scheduling. In \cite{Li-ICL-UAV32}, the minimum SR of ground terminals was maximized by jointly optimizing the UAV flight trajectory, downlink transmission power and user association. To satisfy the legitimate users' content requests and prevent Eve from eavesdropping, the authors in \cite{Wu_ICASSP_UAV165} proposed a latency-minimized transmission scheme, where the UAV trajectory and user association were jointly optimized. 

For multiple Eves, worst-case SR is often considered for robust design, where the Eve with largest wiretap channel capacity is critical. For such a scenario, \cite{Savkin2020wcl} considered a problem of securing UAV communication in the presence of stationary or mobile Eves, where the authors developed a model predictive control (MPC) based navigation scheme and minimized the energy expenditure of the UAV via online 3D trajectory planning. A UAV-aided multicasting system was investigated in \cite{wang2020establishing}, where the UAV disseminated a common file to multiple  users in the presence of multiple Eves. The UAV sequentially visited the locations of virtual BSs to serve the legitimate users in groups. The paper first optimized the virtual BS locations, and then optimized the trajectory of the UAV via traveling salesman problem (TSP) algorithm. 

For the multi-UAV case, inter-UAV interference and collision avoidance constraints should be considered. The paper \cite{liu2020multi17} maximized the minimum SR of users by jointly optimizing user scheduling, UAVs' trajectories and transmit power allocation under the propulsion energy and the inter-UAV interference constraints. In particular, curvature radius constraints were introduced to avoid the sharp swerve of UAVs. \cite{Li2020wcl} investigated a UAV communication network from a cooperation perspective, where two UAVs transmit confidential messages to their specified ground destinations by sharing the same spectrum, and the two UAVs meanwhile act as cooperative jammers for each other. An iterative algorithm was proposed to maximize the system SR by jointly optimizing the trajectory and transmit power of both UAVs.

In some works, location error model was considered, where the exact locations of Eves were unknown. The worst-case SR was usually considered for this case. The authors in\cite{Cai2019-UAV97} investigated secure communication for UAV-OFDMA systems with multiple legitimate users and a malicious Eve located in an uncertain region. To facilitate the secure communication and save the energy consumption, they maximized the SEE by jointly optimizing the transmission power, UAV's trajectory and user scheduling. With a similar distribution model for Eves, the authors in \cite{Cui2018-UAV144} maximized the average worst-case SR by jointly optimizing the robust trajectory and the transmission power. Further considering multiple Eves distributed in a uncertain region, the authors in \cite{Fan2019-UAV143} jointly optimized the UAV's trajectory, transmission power and user scheduling to maximize the SR of the system.

The UAV may act as a secondary transmitter in a cognitive radio (CR) network\cite{Gao_IAccess_UAV166, Zhou_TC_UAV158}. In such CR scenarios, the UAV should be not only away from the Eves to avoid eavesdropping, but also away from the primary users to avoid interference.  \cite{Gao_IAccess_UAV166} investigated PLS in a UAV-aided CR network with imperfect location information of Eves. The average worst-case SR at secondary receiver was maximized by jointly optimizing the robust trajectory and the transmission power of the UAV. By taking into account two practical location estimation error models \cite{Zhou_TC_UAV158}, namely, the bounded location error model and the probabilistic location error model, two problems including worst-case robust ASR maximization and outage-constrained robust ASR maximization were formulated, where the UAV's trajectory and transmission power was robustly optimized.

Some advanced techniques can be incorporated into the UAV systems to improve security performance, such as NOMA, SWIPT, mmWave, and coordinated multiple points (CoMP) transmission\cite{sun2021iot,Hong2019access,wang2020tcom,Tang2020access,Yao2019-UAV61,Yao2020twc}. Considering the limited on-board energy, SWIPT has been adopted to provide energy for UAVs. Secure transmission for mmWave SWIPT UAV networks was investigated in \cite{sun2021iot}.  Considering both beamforming design and 3D antenna  gain, the directional modulation (DM) technique based on random frequency diverse array (RFDA) is adopted to guarantee secrecy. In \cite{Hong2019access}, an efficient resource allocation problem was studied to guarantee the secure communications of the UAV-assisted SWIPT systems with multiple Eves, where the SR was maximized via jointly designing the trajectory and the transmission power. NOMA is seen as a promising technique in UAV networks to improve the spectral efficiency when serving multiple users. A downlink secure UAV-NOMA network was investigated in \cite{wang2020tcom}, where the users are categorized as security-required users (SUs) and quality of service (QoS)-required users (QUs), and the QUs potentially act as internal Eves. The minimum SR among SUs are maximized by jointly optimizing  user scheduling, power allocation, and trajectory design, subject to the QoS requirements of QUs and the mobility constraint of UAV-BS. In \cite{Tang2020access}, a moble UAV-enabled CR-NOMA  system was considered. The 3D UAV trajectory and power allocation were optimized to maximize the worst-case average secrecy sum rate of all secondary users under the interference constraint at primal receivers. Secure UAV communication with CoMP reception was investigated in \cite{Yao2019-UAV61}, where one UAV serves a set of coordinated ground nodes in the presence of several colluding Eves. To facilitate secure communication, they jointly optimized the 3D trajectory and transmit power allocation. In a similar scenario, \cite{Yao2020twc} provided analysis for both non-colluding and colluding Eves.

\textit{\textbf{2) UAV helper :}} In this subsection, we further introduce the trajectory optimization works where UAVs act as helper for PLS, i.e., relay or friendly jammer.
Similarly, we start with a basic four-node model, where a source intends to send confidential message to a legitimate Rx with the help of a UAV relay, in the presence of an Eve.  
Such a basic four-node model was discussed in \cite{Shen-CyberC-UAV136,SunGlobecom2018-UAV42, Xiao-TGCN-UAV44, Wang2018-UAV20}. In \cite{Shen-CyberC-UAV136}, the authors investigated trajectory optimization under constant UAV's transmission power. An effective algorithm by optimizing the increments at each trajectory iteration was proposed, which significantly improved the system's SR. An alternating optimization approach was proposed in \cite{SunGlobecom2018-UAV42} to address the non-convex optimization problem. By jointly optimizing the communication resources, i.e., power and bandwidth allocation, and the UAV's trajectory, the accumulated SR was maximized. Considering a practical channel model, including both LoS and NLoS fadings, the authors in \cite{Xiao-TGCN-UAV44} aimed to strike a trade-off between the SR and the energy consumption of the UAV via maximizing the SEE, in which the communication scheduling, the source/relay power allocation, and UAV's trajectory were jointly optimized. Under practical mobility and information-causality constraints, \cite{Wang2018-UAV20} focused on maximizing the SR via jointly optimizing the mobile relay's trajectory and the source/relay transmission power.

The above mentioned works mainly assumed that the position of  Eve was fixed and known. More practical and complicated assumptions on the Eve distribution were further discussed in \cite{ShenVTC-UAV55, Shensenseors-UAV55, SunICC2019-UAV52}. Assuming that the Eve is uniformly distributed on a line, authors in \cite{ShenVTC-UAV55, Shensenseors-UAV55} investigated the optimal UAV trajectory design to maximize the sum rate, considering the buffer size limit on the UAV relay. The potential Eve's locations were modelled as a disk area on the ground in \cite{SunICC2019-UAV52}, where the authors investigated the robust trajectory and resource allocation design for UAV-aided wireless communications.
  
For the multi-node scenarios, the previous analysis and design methods in general still apply, while more various optimization objectives, variables, and constraints appear according to the more complicated scenario. Considering UAV relaying wireless networks with caching, the authors in \cite{TC-Cheng-UAV4, Cheng-ICC-UAV38} proposed a novel scheme to guarantee secure transmission between data-requiring users and cache-enabled users. By jointly optimizing the trajectory and time scheduling, the minimum SR of the users without cache was maximized. The authors in \cite{Ahmed2019-UAV140} considered the UAV as a mobile relay to assist BS to disseminate information to multiple users in the presence of multiple Eves. The transmission power of BS and UAV, and the UAV trajectory were jointly optimized, to maximize the average worst-case SR. In  \cite{ahmed2020joint}, the authors investigated a UAV relaying network, where multiple adversaries tried to intercept the legitimate link, and their position information was uncertain. Considering both the worst-case SR and UAV propulsion energy consumption, the authors maximized the EE by optimizing the UAV/BS transmit power and UAV's trajectory. A cache-enabled UAV-relaying wireless network was considered in  \cite{ji2020joint21,ji2020joint22}, where UAVs were employed as mobile relays to ferry the information from the macro BS to users, and both UAVs and D2D users were equipped with cache memory. In \cite{ji2020joint21}, user association, UAV scheduling, transmission power, and UAV trajectory  were jointly optimized to maximize the minimum SR among users, while in \cite{ji2020joint22}, the cache placement was jointly considered with trajectory optimization. The study in \cite{Chi-Nguyen2018-UAV137} considered a CR system, where both information-causality and interference constraints were considered in the joint power and trajectory optimization.

Friendly jammer is another important role for UAV helpers to enhance the PLS. In general, the trajectory design should make the UAV jammer close to Eves while being away from the legitimate users. We first introduce the four-node model where a UAV jammer transmits jamming signals against eavesdropping in a ground wiretap system. By jointly controlling the transmission power of both the transmitter and the UAV jammer, the ASR was maximized in \cite{Li_ICL_UA138}.  By assuming an ellipse model for Eve's unknown location, the authors in \cite{Nnamani2020access} investigated UAV-aided jamming  for secure ground communications with the target of  average secrecy rate maximization. Under the uncertainties of the nodes' locations, a robust joint design problem of optimizing the UAV's jamming power and trajectory was formulated to maximize the average secrecy rate \cite{Roh2019milcom}. The authors in \cite{nguyen2020uav} investigated the scenario where a UAV secured an underlay cognitive radio network (CRN) as a friendly jammer, and interfered the Eve who attempts to decode the confidential message from the secondary transmitter (ST). The transmit power and UAV’s 3D trajectory were jointly optimized to maximize the average achievable secrecy rate of the secondary system subject to the primary receiver’s interference power constraint. 

Some works further considered the composite case, that one UAV can may act as jammer, communication node, or a relay simultaneously \cite{Xu2021tcom, Duo2020tvt, li2020uav, zhou2020uav}. Considering the limited on-board energy of UAVs, the power allocation between communication and jamming should be carefully designed. A UAV-enabled secure communication system was investigated in \cite{Xu2021tcom}, where the UAV as a legitimate transmitter employed power splitting for transmitting confidential messages and artificial noise. By jointing optimizing the UAVs' trajectory, the transmission power and the power splitting ratios, the ASR was maximized. The authors in \cite{Duo2020tvt} investigated EE maximization for FD UAV secrecy communication, where the UAV sent jamming signals to potential ground Eve, while receiving the confidential messages. The paper \cite{li2020uav} considered a single UAV-enabled secure data dissemination system where the UAV secretly communicate with multiple ground users by TDMA protocol. AN was introduced to prevent confidential signal intended to scheduled users being overheard by unscheduled ones. To guarantee fairness among all ground users, it maximized the minimum ASR by jointly optimizing the UAV trajectory, transmit power splitting ratio and the user scheduling.  The authors in \cite{zhou2020uav} investigated a UAV-enabled confidential data collection problem, and derived the reliability outage probability (ROP) and SOP. They maximized the minimum ASR by optimizing the UAV trajectory, AN transmit power, transmission rates and the sensor node scheduling, under the ROP and SOP constraints.

Cooperation between a UAV jammer and a UAV communication node can enhance the PLS, and their trajectories can be designed jointly. This is known as the dual-UAV scenario, which has addressed increasing research attentions recently. In this case, the inter-UAV distance constraint should be considered when designing their trajectories. A basic four-node model was considered in \cite{Zhong_ICL_UAV24, Mamaghani2020tvt, li2018mobile}, where two UAVs acted as Tx and jammer, respectively. The authors in \cite{Zhong_ICL_UAV24} adaptively adjusted the UAVs' trajectories to maximize the ASR, where the constraints of the transmission/jamming power was considered with  partially-known Eve's location. In \cite{Mamaghani2020tvt}, the authors considered two jamming schemes, namely friendly UAV jamming and Gaussian jamming. With partial knowledge of Eve's location, an optimization scheme with robust trajectory design and communications resource allocation was proposed to maximize the ASR. In \cite{li2018mobile}, the authors maximized the SR via trajectory design and power control of both source UAV and jammer UAV.  The work in \cite{Nguyen_CCNC_UAV85} studied UAV jamming in a CR network. Under a given interference threshold at the primary receiver, the SR was maximized by jointly optimizing the UAV's trajectory and the transmission power.

UAV jammer/Tx cooperation was further investigated for the multi-user or multi-Eve cases in \cite{Lee_TVT_UAV28, Sun2019sensors, Li_Access_UAV37, Cai_JASC_UAV9, Cai_ICC_UAV59, Zhou_TVT_UAV10, Cai2020tcom}, where user scheduling becomes another dimension to be optimized, and worst-case SR is usually considered in the design. To guarantee fairness for multiple users, joint trajectory, transmit power, and user scheduling optimization was conducted to maximize the minimum SR in \cite{Lee_TVT_UAV28}. In \cite{Sun2019sensors}, to maximize the worst-case ASR, the two UAVs' trajectories and the BS’s/UAV jammer's transmit/jamming power were jointly optimized. The paper \cite{Li_Access_UAV37} considered multiple location-unknown Eves. Joint trajectory design and the transmission power allocation for both UAVs were performed to maximized the worst-case SR. The authors in \cite{Cai_JASC_UAV9, Cai_ICC_UAV59} developed a new algorithm based on the penality concave-convex procedure technique to handle secure communications for a dual-UAV enabled networks. Joint optimization of UAVs' trajectories and user scheduling was  performed to maximize the minimum ASR. In \cite{Zhou_TVT_UAV10}, the authors considered multiple users and multiple Eves, and maximized the minimum ASR over all users by jointly optimizing the UAV trajectory and transmission power. The paper \cite{Cai2020tcom} considered an energy-efficient secure UAV-OFDMA communication system, where a UAV Tx with the assistance of a multiple-antenna UAV jammer provides secure communications for multiple users in the existence of multiple Eves. To maximize the EE, the trajectory, resource allocation, and jamming policy were jointly designed. For the multi-jammer case, the authors in \cite{zhang2020uav32,zhang2020multi32} considered the cooperation between the UAV jammers and UAV Tx, where the UAV jammers send AN signals to ground Eves by 3D beamforming. A multi-agent deep reinforcement learning approach was proposed to maximize the secure capacity by jointly optimizing the trajectory of UAVs, the transmit power and the jamming power. The paper \cite{Hua_TVT_UAV7} further considered a cooperative scenario with multiple source UAVs and jammer UAVs. The SEE was maximized by jointly optimizing the UAV trajectory, transmit power and user scheduling. 
  
UAV jammer has also been integrated into UAV relaying systems \cite{XU2021871, Miao_Access_UAV156, Zhang2019infocom, Chi-Nguyen2018-UAV137}. In \cite{XU2021871}, for such dual-UAV system with one UAV relay and one UAV jammer, the authors proposed an iterative algorithm via optimizing the trajectories and transmit power for both UAVs to maximize the SR. In \cite{Miao_Access_UAV156}, the constraints of flight trajectory and power on both UAVs, as well as the information-causality constraint on the relaying UAV, were jointly considered in the optimization to maximize the SR. The paper \cite{Zhang2019infocom} considered a multi-UAV communication network, where a novel joint friendly jamming and bandwidth allocation scheme was developed.

\textit{\textbf{3) UAV Attacker and Covert Communications:}}  
As compared to the UAV roles described above, trajectory optimization works in the presense of UAV attackers, and that for covert communications, are relative less. We introduce several representative works in this one subsection.
  
The UAV attacker can be more harmful compared with a ground attacker, because of the high probability of LoS propagation, as well as its high mobility. The authors in \cite{WuICCC-UAV145} considered UAV Eve in a UAV millimeter wave communication system. The trajectory of the legitimate UAV Tx was designed to avoid UAV Eve appearing in a cone region below it. Considering UAV jamming attack, the authors in \cite{Wang_WCSPUA198} investigated the 3D deployment and trajectory planning problem by dynamically adjusting the UAV’s trajectory to maximize the sum throughput with jamming. 

For covert communications, a legitimate UAV communication node should try to approach the other legitimate nodes during transmission, so as to reduce the communication power and hence reduce the probability of being detected. Besides, AN can be employed for better covertness performance. In \cite{Zhou2019globec,zhou2019uav}, a convert communication problem was formulated, where a UAV collects data covertly from ground users and meanwhile generates AN to ensure the scheduled user's transmission not being detected by unscheduled users. The UAV's trajectory, AN power, and user scheduling were jointly optimized to maximize the minimum transmission rate, subject to covertness constraints. Convert communications in a UAV network was investigated in \cite{Zhou_TSP19}, where a UAV Tx intends to communicate with a ground receiver covertly, while a ground warden wanted to detect the UAV's transmission.  The UAV's trajectory and transmit power were jointly optimized to maximize the average covert transmission rate, subject to both transmission outage and covertness constraints.

    \vspace{-3mm}
\subsection{Summary of This Section}

We summarize what have been learnt in this section:

\begin{itemize}
\item For mobile UAV deployment, the PLS design is usually conducted in a communication process-oriented manner. In other words, the optimization now needs to span the entire period that communication happens. The trajectory of the UAVs, either being random, fixed, or to be optimized, introduces new constraints or objectives in the optimization for different PLS scenarios.

\item When the UAV trajectory is not controllable by the communication designer and considered as a fixed condition, it is usually reflected in the constraints of optimization. For example, a certain fixed trajectory will directly impact the time-domain power allocation, or the information-causality when UAV acts as a relay. 

\item Accompanying mobility model can be seen as a special case of fixed trajectory. A UAV Tx or relay may follow the trajectory of ground user to provide better service. For wireless surveillance, a UAV monitor may follow the trajectory of the suspicious nodes for better detection. In practice, the accompanying trajectory can be realized based on signal strength or vision, and the accompanying strategy can be jointly designed with the communication strategies, to enhance the PLS performance.

\item Trajectory optimization contributes a major research direction in the UAV-related PLS works. The optimization methodology that is adopted for different scenarios can be similar, while the problem formulation can be largely different. Realistic constraints, such as collision avoidance inter-UAV distance, UAV propulsion energy, partial knowledge of Eve's CSI, etc., are usually introduced to make the problem formulation more practical. 
\end{itemize}

\section{Potential Future Directions and Challenges}

\subsection{MIMO, RIS, and UAV Swarm-enabled CoMP}

Due to the limited payload, a large amount of existing UAV-PLS works consider only single antenna onboard. To further enhance the communication performance, it is desired to further exploit the spatial dimension in the A2G channel. Thanks to the latest developments in radio technologies, it becomes possible to pack more antenna/reflection elements in a compact sized array. This has promoted the research on MIMO/RIS-aided UAV communications, where the PLS design problems need to be re-formulated. In particular, the design or MIMO precoding, AN matrix, or RIS phase shifters, needs to be considered in addition to the conventional optimization dimensions. Though there has already emerged some pioneer works on this research direction (some have been introduced in previous sections), the number of corresponding researches is still far less than its counterpart, where single-antenna UAV platforms were considered. Many practical scenarios are yet to be extended to the MIMO/RIS case.

On the other hand, when MIMO/RIS-based UAV platforms are not available, single antenna UAVs can form a swarm-based virtual MIMO systems, performing CoMP transmission to ground users. This is becoming possible as the swarm coordination techniques are becoming mature. For such scenario, new PLS problems arise, e.g., malicious UAV detection in a swarm. A malicious UAV is able to follow the legitimate UAVs  to perform eavesdropping/jamming attacks more effectively, meanwhile, the attack could be more harmful due to the favorable A2A channels. The detection, surveillance, avoidance, and possible jamming of/on potential malicious UAVs in a swarm is an important issue that needs to be considered. Furthermore, possible attack on the control signals also becomes a critical issue \cite{Abughalwa_CL20,Huang_TVT18}, as the UAV swarm-enabled communication highly relies on reliable signalling channels among them.

\vspace{-2mm}
\subsection{Integration with Satellite Networks}

Both satellites and UAVs are important candidates for non-terrestrial network beyond 5G, which can be utilized to cover the blind zones of terrestrial cellular networks. In the current developing stage, both satellites and UAVs have their drawbacks, individually, in terms of latency or endurance. This problem can be alleviated by proper coordination between satellites and UAVs to form an integrated satellite-UAV network. As satellite transmission is more open than UAVs, this may lead to more serious secrecy problem, which should be carefully addressed in the PLS design. In such a scenario, the UAV may act as a helper to safeguard satellite communications thanks to its agile mobility \cite{Bankey_UAV168}. The heterogeneity of such a network, especially the signalling and data exchange overhead for coordination, as well as joint resource allocation, should be carefully modelled in the optimization.

\vspace{-2mm}
\subsection{More Powerful Helpers: Caching and MEC}

In addition to the conventional roles of helper, i.e., relay and friendly jammer as described above, UAV can be deployed as more powerful helpers, who can offload the user-desired data from the BS (caching), or it can offload computation tasks from the end-users (MEC). Flexible deployment of UAV is promising to make either caching or MEC more effective, however, secrecy issues need to be addressed. Possible impacts of caching on communication secrecy has been investigated from the following perspectives: The cache hit rate of a user may affect its communication link selection. That is, if the user cannot find its desired content in the UAV cache, it turns to the source node for the data. In the presence of Eve, different communication link selection subsequently results in different secrecy performance, as has been discussed in \cite{Li_PhyCom19}. On the other hand, when a user can acquire its desired content in UAV cache, the source nodes can be idle for transmission and therefore can be utilized for friendly jamming. This will help improve the secrecy performance, as has been investigated in \cite{Zhao-TC2018-UAV1,Cheng-ICNC-2018-UAV8}. In general, the feature of UAV caching is reflected by its deployment flexibility: For example, the number of UAVs required for secure transmission has been discussed in \cite{Zhao-TC2018-UAV1}. As for MEC-empowered UAVs, computing would reduce the amount of data for further transmissions, thus provides better secrecy. However, this requires joint resource orchestration of both computing and communications. There are still lots of works to be done in these research areas.

\vspace{-2mm}
\subsection{Machine Learning Approaches}

When multiple UAVs are adopted in a partially unknown environment, the optimization of PLS would become much more complex. On the one hand, the UAV swarm control and communications design, as well as caching and computing if integrated, should be fully taken into account. However, some parts of which are hard to model due to interdisciplinarity. On the other hand, the optimization problem is usually non-convex with high-dimensional variables, it is challenging to design  low-complexity methods to solve the problem with affordable performance degradation. The fast developing machine learning techniques provide promising and powerful tools for this tricky scenario. It is possible to derive a hybrid model and data-driven method to characterize the complex system, and use hybrid offline and online approaches to optimize the secrecy performance. For example, the UAV trajectory design can be promoted by reinforcement learning. Nevertheless, using machine learning approaches may raise new secrecy problem, in e.g., data gathering or parameter updating. Recently, a number of works have emerged in the area of machine learning-aided secure UAV communications \cite{Xiao2018tvt,xiao2018tvtCentric,Gao2020tcom,
Challita2019twc,Min2018vtcspring,Hoang2020wcl,
Liu2019wcsp,khadem2020efficient,Bao2020wcl}. However, there still exist much more scenarios that machine learning may find its application, and a systematic design framework is still needed for the learning-aided secure UAV communications.

\section{Conclusions}
The article surveyed important issues of UAV communications from the perspective of PLS. As necessary preliminary of the survey, we first introduced typical application scenarios with static/mobile UAV deployment, the unique channel and location distribution models in UAV communications, as well as typical roles of UAV for PLS. After that, we provided an exhaustive review of the state-of-art research achievements on the UAV-PLS, which were categorized into two cases: the UAV is deployed at static positions, and the UAV is moving during the communication process following certain trajectory. Respectively for each case, we have summarized the commonly adopted methodology in the analysis and design, and discussed important literatures in detail. Finally, we have further shown several promising research directions and possible challenges for future researches.

\appendices



\end{document}